\documentclass[journal,draftclsnofoot,onecolumn,12pt]{IEEEtran}
\pdfoutput=1
\usepackage[colorlinks=true,linkcolor=blue,urlcolor=black,bookmarksopen=true]{hyperref}
\usepackage{amsthm,amssymb,graphicx,multirow,amsmath,color,amsfonts}
\usepackage[update,prepend]{epstopdf}
\usepackage[noadjust]{cite}
\usepackage[latin1]{inputenc}
\usepackage{tikz}
\usetikzlibrary{arrows,calc}		
\usepackage{bbm} 
\usepackage{pdfpages}
\usepackage{tabulary}
\usepackage{multirow}
\usepackage{comment}
\usepackage{mathtools}
\usepackage{bigints}
\usepackage{adjustbox}
\usepackage{setspace}

\def\nb0{{\mathbf{0}}}
\def\nb1{{\mathbf{1}}}

\def\nbbE{{\mathbb{E}}}

\def\nbbR{{\mathbb{R}}}

\def\nrmx{{\rm x}}
\def\nrmy{{\rm y}}

\newtheorem{lemma}{Lemma}
\newtheorem{thm}{Theorem}
\newtheorem{definition}{Definition}

\newtheorem{remark}{Remark}
\newtheorem{assumption}{Assumption}
	
\def\argmax{\operatorname{arg~max}}

\def\P{\mathbb{P}}
\def\pc{\mathtt{P_c}}
\def\rc{\mathtt{R_c}}   

\def\erfc{\operatorname{erfc}}

\def\T{\beta}							

\def\sir{\mathtt{SIR}}

\def\calC{\mathcal{C}}

\def\calE{\mathcal{E}}

\def\calL{\mathcal{L}}

\def\calR{\mathcal{R}}

\def\calV{\mathcal{V}}

\def\calX{\mathcal{X}}

\def\calZ{\mathcal{Z}}

\allowdisplaybreaks 

\let\subparagraph\relax 

\usepackage[compact]{titlesec}
\titlespacing{\section}{0pt}{*.5}{*0.5}
\titlespacing{\subsection}{0pt}{*.5}{*.5}

\begin{document}
\graphicspath{{./Figures/}}
\title{Coverage and Rate Analysis of Downlink Cellular Vehicle-to-Everything (C-V2X) Communication }
\author{
Vishnu Vardhan Chetlur and Harpreet S. Dhillon
\thanks{The authors are with Wireless@VT, Department of ECE, Virginia Tech, Blacksburg, VA (email: \{vishnucr, hdhillon\}@vt.edu). The support of the US NSF (Grant IIS-1633363) is gratefully acknowledged. \hfill
\hfill Manuscript last updated: \today.}
}
\maketitle
\begin{abstract}
In this paper, we present the downlink coverage and rate analysis of a cellular vehicle-to-everything (C-V2X) communication network where the locations of vehicular nodes and road side units (RSUs) are modeled as Cox processes driven by a Poisson line process (PLP) and the locations of cellular macro base stations (MBSs) are modeled as a 2D Poisson point process (PPP). Assuming a fixed selection bias and maximum average received power based association, we compute the probability with which a {\em typical receiver}, an arbitrarily chosen receiving node, connects to a vehicular node or an RSU and a cellular MBS. For this setup, we derive the signal-to-interference ratio (SIR)-based coverage probability of the typical receiver. One of the key challenges in the computation of coverage probability stems from the inclusion of shadowing effects. As the standard procedure of interpreting the shadowing effects as random displacement of the location of nodes is not directly applicable to the Cox process, we propose an approximation of the spatial model inspired by the asymptotic behavior of the Cox process. Using this asymptotic characterization, we derive the coverage probability in terms of the Laplace transform of interference power distribution. Further, we compute the downlink rate coverage of the typical receiver by characterizing the load on the serving vehicular nodes or RSUs and serving MBSs. We also provide several key design insights by studying the trends in the coverage probability and rate coverage as a function of network parameters. We observe that the improvement in rate coverage obtained by increasing the density of MBSs can be equivalently achieved by tuning the selection bias appropriately without the need to deploy additional MBSs.
\end{abstract}
\section{Introduction} \label{sec:intro}
Vehicular communication networks are essential to the development of intelligent transportation systems (ITS) and improving road safety \cite{survey, traffic_safety, vc_its}. As the in-vehicle sensors can assess only their immediate environment, some information still needs to be communicated from external sources to the vehicle to assist the driver in making critical decisions in advance \cite{usecases}. Vehicle-to-vehicle (V2V) communication enables the vehicular nodes to share information with each other without network assistance. Vehicle-to-infrastructure (V2I) and vehicle-to-network (V2N) communications can support a wide range of applications from basic safety messages and infotainment to autonomous driving. The underlying technology that enables all these services through long term evolution (LTE) communication is referred to as cellular vehicle-to-everything (C-V2X) and has been standardized by the third generation partnership project (3GPP) as part of Release 14 \cite{3gpp}. The key technical characteristics of C-V2X include low latency, network independence, and support for high speed vehicular use. Some preliminary system-level simulations are considered in \cite{3gpp} to study the performance of this network under different scenarios. Since simulation-based design approaches are often time consuming and may not be scalable for large number of simulation parameters, it is important to develop complementary analytical approaches to gain design insights and benchmark simulators. For this purpose, tools from stochastic geometry are of particular interest, where the idea is to endow appropriate distributions to the locations of different network entities and then use properties of these distributions to characterize network performance. From the perspective of vehicular networks, the spatial model that is gaining popularity is the so-called {\em Cox process}, where the spatial layout of roads is modeled by a Poisson line process (PLP) and the locations of vehicular nodes and road side units (RSUs) on each line (road) are modeled by a 1D Poisson point process (PPP) \cite{vishnuJ2, bacc_plp}. While this spatial model has been employed in a few works in the literature, the effect of shadowing on the performance of the vehicular network has not been considered. Also, the characterization of key performance metrics such as \textit{rate coverage}, which requires the computation of load on the cellular macro base stations (MBSs), vehicular nodes, and RSUs, is still an open problem. In this paper, we close this knowledge gap and present the coverage and rate analysis for a C-V2X network in the presence of shadowing. 
\subsection{Related Work}
Most of the earlier works in the stochastic geometry literature pertaining to vehicular communications were motivated by vehicular ad hoc networks (VANETs) based on dedicated short range communication (DSRC) standard \cite{mit, busanelli, mabiala, Ni, Steinmetz}. Also, the spatial models considered in these works were limited to a single road or an intersection of two roads. That said, there have been some works that consider more sophisticated models, primarily the Cox process, where the spatial layout of roads were modeled by a PLP and the locations of nodes on each road were modeled as a 1D PPP \cite{bacc_plp, vishnuL1, multihop,robert,  morlot, bartek, dim5g,vishnuJ2, bacc_letter}. The idea of using PLP to model road systems was first proposed in \cite{bacc_plp} to study the handover rate of vehicles moving across cells. This model was further explored in \cite{volker}, where the distribution of the distances between the nodes located on a PLP was studied. 
In \cite{vishnuL1}, the authors have presented the success probability of a typical link in a VANET modeled as a Cox process driven by a PLP. The routing performance of a VANET for a linear multi-hop relay was studied in \cite{multihop}.
In \cite{robert}, the authors have presented the downlink coverage probability of a typical receiver in an urban setup where the locations of base stations operating at millimeter wave frequencies were modeled as Cox process driven by a Manhattan Poisson line process (MPLP). The uplink coverage probability for a setup where the typical receiver is chosen from a 2D PPP and the locations of transmitting nodes form a Cox process was presented in \cite{morlot}. In \cite{vishnuJ2, vishnuICC}, the authors have derived the downlink coverage probability of a typical receiver for nearest neighbor connectivity in a vehicular network where the transmitter and receiver nodes are modeled by Cox processes driven by the same PLP. While these are key steps towards understanding the behavior of vehicular networks, the analysis presented in these works is often limited to V2V and V2I communications as they do not include cellular base stations in their setup.

Recently, there have been a few works in which cellular-assisted vehicular networks were considered \cite{baccchoi, sguha, Sial}. A comprehensive coverage analysis for two different types of users {(planar and vehicular users)} was provided in \cite{baccchoi}, where the locations of transmitting vehicular nodes were modeled by the Cox process and the locations of cellular base stations were modeled by 2D PPP. In \cite{Sial}, the authors have attempted to derive the uplink coverage probability for C-V2X communication. However, the effects of shadowing were ignored in these works because of the technical complications arising from the doubly stochastic nature of this Cox process. So, this paper makes two key contributions. First, we present the coverage analysis of a C-V2X network including the shadowing effects in our model. Second, we present the rate coverage analysis for a C-V2X network. 
More details of our contribution are provided next. 
\subsection{Contributions}
We model the locations of vehicular nodes and RSUs by Cox processes driven by a PLP and the locations of cellular MBSs by a homogeneous 2D PPP. We consider Nakagami-$m$ fading to model the effects of small-scale fading. We consider standard power-law path-loss model and log-normal shadowing to model large-scale fading effects. We further introduce selection bias to balance the load between MBSs, RSUs and vehicular nodes across the network. Assuming that a typical receiver, which is an arbitrarily chosen receiver node, connects to the node that yields the highest average biased received power among the contending nodes, we derive the signal-to-interference ratio (SIR)-based coverage probability and rate coverage. We also offer useful design insights by studying the trends in the performance as a function of key network parameters. More technical details about the analysis are provided next.

{\em Asymptotic behavior of the Cox process.} We first provide the expression for the void probability of a Cox process driven by a PLP for any planar Borel set. We then rigorously show that this expression asymptotically converges to that of the void probability of a homogeneous 2D PPP with the same average density of points, thereby proving the convergence of the Cox process to a 2D PPP by Choquet's theorem \cite{haenggi}.

{\em Coverage probability.} The key technical challenge in the characterization of SIR-based coverage probability stems from the inclusion of shadowing effects. Traditionally, in wireless networks where the locations of nodes were modeled as a PPP, the effect of shadowing could be conveniently interpreted as a random displacement of the location of the nodes in the computation of the desired signal power and interference power at the receiver \cite{Dhi_letter}. However, this method is not applicable to the Cox process as the random displacement of each point on a line would disrupt the collinearity of the points, thereby making it difficult to exactly characterize the SIR at the typical receiver. Therefore, inspired by the asymptotic behavior of the Cox process, we propose a tractable and accurate approximation of the spatial model. Using this asymptotic characterization, we obtain the expression for coverage probability in terms of the Laplace transform of the interference power distribution. Given its nature, the proposed model could enable similar analyses that may not otherwise be possible due to the doubly stochastic nature of a Cox process driven by a PLP. 

{\em Rate Coverage.} In order to determine the rate coverage of the typical receiver, one of the important components is the distribution of load on the serving node of the typical receiver. As the load on the serving node is proportional to the size of its coverage region, we characterize the length of the coverage region of the serving vehicular node or RSU. We also determine the mean load on the serving MBS. Using these results, along with the distribution of SIR, we completely characterize the rate coverage of the typical receiver.

{\em Design Insights.} Using the analytical results, we study the impact of node densities and selection biases on the coverage probability and rate coverage of the typical receiver. As expected, we observe that the rate coverage can be improved by increasing the node density of MBSs. However, since the deployment of additional MBSs is costly (and may not always be possible because of challenges in site acquisitions), {our results concretely demonstrate that we can alternately achieve similar performance gains by adjusting bias factors appropriately}. We also observe that an increase in the density of RSUs has a contrasting effect on coverage probability and rate coverage. While a denser deployment of RSUs decreases coverage probability due to an increase in interference power, it also reduces the load on the RSUs thereby improving the rate coverage. Hence, one must consider such trade-offs between these two metrics in the design of the network.

\section{Mathematical Preliminary: Asymptotic Behavior of Cox Process}\label{sec:prelim}
In this section, we will explore a key property of the Cox process driven by a PLP which will be useful in developing a tractable way to characterize the interference distribution in our analysis later. Specifically, we will analyze the asymptotic behavior of a Cox process driven by a PLP for extreme values of line and point densities. As basic knowledge of PLP and its properties would be useful in understanding this analysis, we will first give a brief introduction to the topic. A detailed account on the theory of line processes can be found in \cite{stoyan}.  

\begin{definition}{\rm  (Poisson line process).}
	 An undirected line $L$ in $\nbbR^2$ can be uniquely parameterized by its perpendicular distance from the origin $\rho$ and the angle $\theta$ subtended by the perpendicular dropped onto the line w.r.t. the positive $x$-axis. So, each line in $\nbbR^2$ can be uniquely mapped to a point with coordinates $(\rho, \theta)$ on the cylindrical surface $\calC \equiv \nbbR^+ \times [0, 2 \pi)$, which is termed as the \textit{representation space}. A random collection of lines that is generated by a PPP in $\calC$ is called a PLP. 
\end{definition}
\begin{definition} {\rm (Line density).}
	The line density of a line process is defined as the mean line length per unit area.
\end{definition}
\begin{definition} {\rm (Motion-invariance).}
A line process is said to be motion-invariant if the distribution of lines is invariant to translation and rotation.
\end{definition}
%

We will now discuss the construction of the Cox process. First, we consider a motion-invariant PLP $\Psi_\ell$ with line density $\mu_\ell$. We denote the corresponding 2D PPP in $\calC$ by $\Psi_{\calC}$ with density $\lambda_\ell$. As shown in \cite{stoyan}, the relation between $\lambda_\ell$ and $\mu_\ell$ is given by $\lambda_\ell = \frac{\mu_\ell}{\pi}$. We populate points on the lines of $\Psi_\ell$ such that they form a 1D PPP with density $\lambda_p$ on each line. Thus, we obtain a Cox process $\Psi_a$ driven by the PLP $\Psi_\ell$. We begin our analysis by deriving the void probability of the Cox process in the following lemma. 

\begin{lemma}\label{lem:void}
	The void probability of the Cox process $\Psi_a$ is given by
	\begin{align}\label{eq:p0}
	\P\left(N_p (A) = 0\right) = \exp \left[- \lambda_\ell  \int_0^{2\pi} \int_{\nbbR^+} \left[1 - \exp\left(  - \lambda_p \nu_1 \left(L_{(\rho, \theta)} \cap A\right) \right) \right] {\rm d} \rho  {\rm d} \theta \right]  ,
	\end{align}
	where $N_p(\cdot)$ denotes the number of points, $A \subset \nbbR^2$ is a planar Borel set, and $\nu_1(\cdot)$ is the 1D Lebesgue measure. 
\end{lemma}\vspace{-.5em}
\begin{IEEEproof}
	See Appendix \ref{app:void}.
\end{IEEEproof}


Using Choquet's theorem \cite{haenggi}, we will now show that the Cox process $\Psi_a$ asymptotically converges to a homogeneous 2D PPP in the following theorem.

\begin{thm}\label{thm:convergencetoPPP}
	As the line density approaches infinity ($\lambda_\ell \to \infty$) and the density of points on each line tends to zero ($\lambda_p \to 0$) while the overall density of points (average number of points per unit area) remains unchanged, the Cox process $\Psi_a$ converges to that of a 2D PPP with the same average node density $\pi \lambda_\ell \lambda_p$. 
\end{thm}\vspace{-.5em}
\begin{IEEEproof}
	See Appendix \ref{app:convergence}.
\end{IEEEproof}

\section{System Model}\label{sec:systemmodel}
\subsection{Spatial Modeling of Wireless Nodes}
\begin{figure}
	\centering
	\includegraphics[width=.5\textwidth]{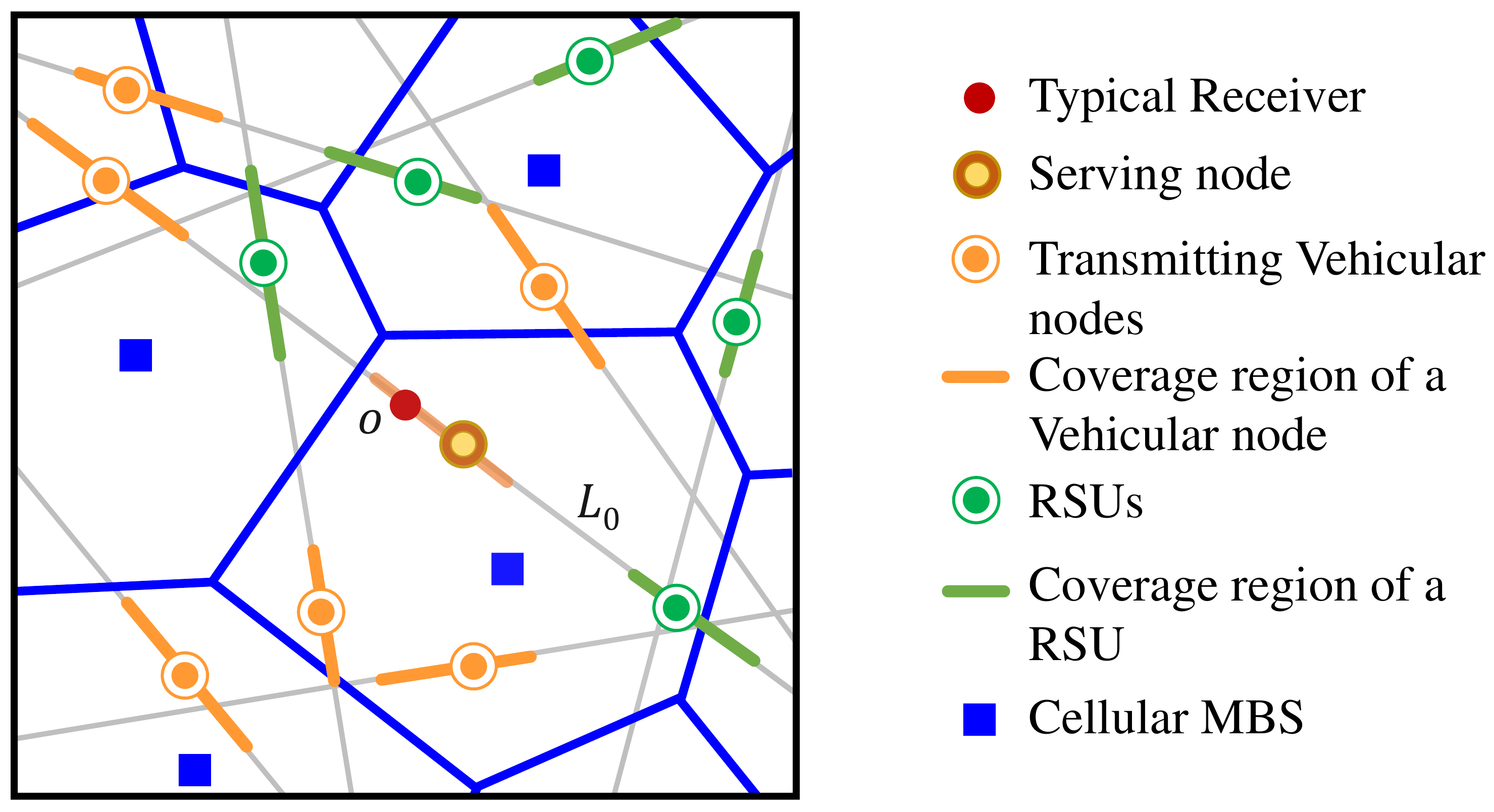}
	\caption{Illustration of the system model.}
	\label{fig:sysmod1}
\end{figure}

We consider a vehicular network consisting of vehicular nodes, RSUs, and cellular MBSs, as illustrated in Fig. \ref{fig:sysmod1}. Since the locations of vehicles and RSUs are restricted to roadways, we first model the irregular spatial layout of roads by a PLP $\Phi_l$ with line density $\mu_l$. We denote the corresponding 2D PPP in the representation space $\calC$ by $\Phi_\calC$ with density $\lambda_l$. 
We model the locations of transmitting and receiving vehicular nodes on each line $L_j \in \Phi_l$ by independent and homogeneous 1D PPPs $\Xi_{L_j}^{(V)}$ and $\Omega_{L_j}^{(R)}$ with densities $\lambda_v$ and $\lambda_{r}$, respectively. Further, we model the locations of RSUs on each line $L_j$ by a 1D PPP $\Xi_{L_j}^{(U)}$ with density $\lambda_u$.
Thus, the locations of receiving vehicular nodes, transmitting vehicular nodes and RSUs form Cox processes $\Phi_r \equiv \{\Omega_{L_j}^{(R)}\}_{L_j \in \Phi_l}$, $\Phi_v \equiv \{\Xi_{L_j}^{(V)}\}_{L_j \in \Phi_l}$ and $\Phi_u \equiv \{\Xi_{L_j}^{(U)}\}_{L_j \in \Phi_l}$, respectively. Note that $\Phi_r$, $\Phi_v$, and $\Phi_u$ are stationary \cite{morlot}. The locations of cellular MBSs are modeled by a 2D PPP $\Phi_c$ with density $\lambda_c$. 

Our goal is to characterize the SIR-based coverage probability and rate coverage of a typical receiver, which is an arbitrarily chosen receiving vehicular node from the point process $\Phi_r$. For analytical simplicity, we translate the origin $o \equiv (0,0)$ to the location of the typical receiver. Since the typical receiver must be located on a line, we now have a line $L_0$ passing through the origin. Thus, under Palm probability of the receiver point process, the resulting line process is $\Phi_{l_0} \equiv \Phi_{l} \cup {L_0}$. This result simply follows from the application of Slivnyak's theorem to the line process $\Phi_l$ or equivalently to the corresponding 2D PPP $\Phi_{\calC}$ in the representation space $\calC$. Thus, the point process of receiving vehicular nodes $\Phi_{r_0} \equiv \Phi_r \cup \Omega_{L_0}^{(R)} \cup \{o\}$ is the superposition of $\Phi_r$, an independent 1D PPP with density $\lambda_r$ on line $L_0$ and an atom at the origin. Consequently, under Palm probability of $\Phi_{r_0}$, the point process of transmitting vehicular nodes $\Phi_{v_0} \equiv \Phi_v \cup \Xi_{L_0}^{(V)}$ is the superposition of the point process $\Phi_v$ and an independent 1D PPP with density $\lambda_v$ on line $L_0$ \cite{vishnuJ2, morlot}. Similarly, under Palm probability of $\Phi_{r_0}$, the point process of RSUs $\Phi_{u_0} \equiv \Phi_u \cup \Psi_{L_0}$ is also the superposition of $\Phi_u$ and an independent 1D PPP $\Psi_{L_0}$ on $L_0$ with density $\lambda_u$. The line $L_0$ will henceforth be referred to as the \textit{typical line}. 

\subsection{Propagation Model}
We assume that all the MBSs have the same transmit powers $P_c$. {Further, we assume that all the vehicular nodes and RSUs have the same transmit power $P_u$. Since the vehicular nodes and RSUs predominantly communicate with the nodes on the road on which they are located, we assume that vehicular nodes and RSUs employ transmit beamforming to maximize the signal power at the receiver nodes. Based on the directional antennas considered in \cite{antenna_meas}, we assume a beam pattern in which the main lobes are aligned along the road on which these nodes are located, as shown in Fig. \ref{fig:beam_pattern_veh}. We assume a sectorized antenna pattern with main lobe and side lobe gains denoted by $G_u$ and $g_u$, respectively.
We assume that the cellular MBSs also use transmit beamforming with main lobe and side lobe gains denoted by $G_c$ and $g_c$, respectively, as shown in Fig. \ref{fig:beam_pattern_mbs}. Unlike vehicular nodes and RSUs with fixed beam patterns, MBSs scan the space for the best beam alignment during the association process. Therefore, the main lobe of the serving MBS is directed towards the typical receiver, i.e., the beamforming gain from the serving MBS is always $G_c$. However, the main lobes of other MBSs may not necessarily be in the direction of the typical receiver. So, we assume that the main lobe of an interfering MBS is directed towards the typical receiver with a probability $q_c$, which depends on the main lobe beamwidth. Thus, the beamforming gain from an interfering MBS is $G_c$ with a probability $q_c$ and $g_c$ with probability $(1-q_c)$.
As the typical receiver may not have prior knowledge of the direction of arrival of the desired signal, we assume an omni-directional antenna at the typical receiver.
\begin{figure}
	\centering
\begin{minipage}{.38\textwidth}
	\centering
	\includegraphics[width=.7\textwidth]{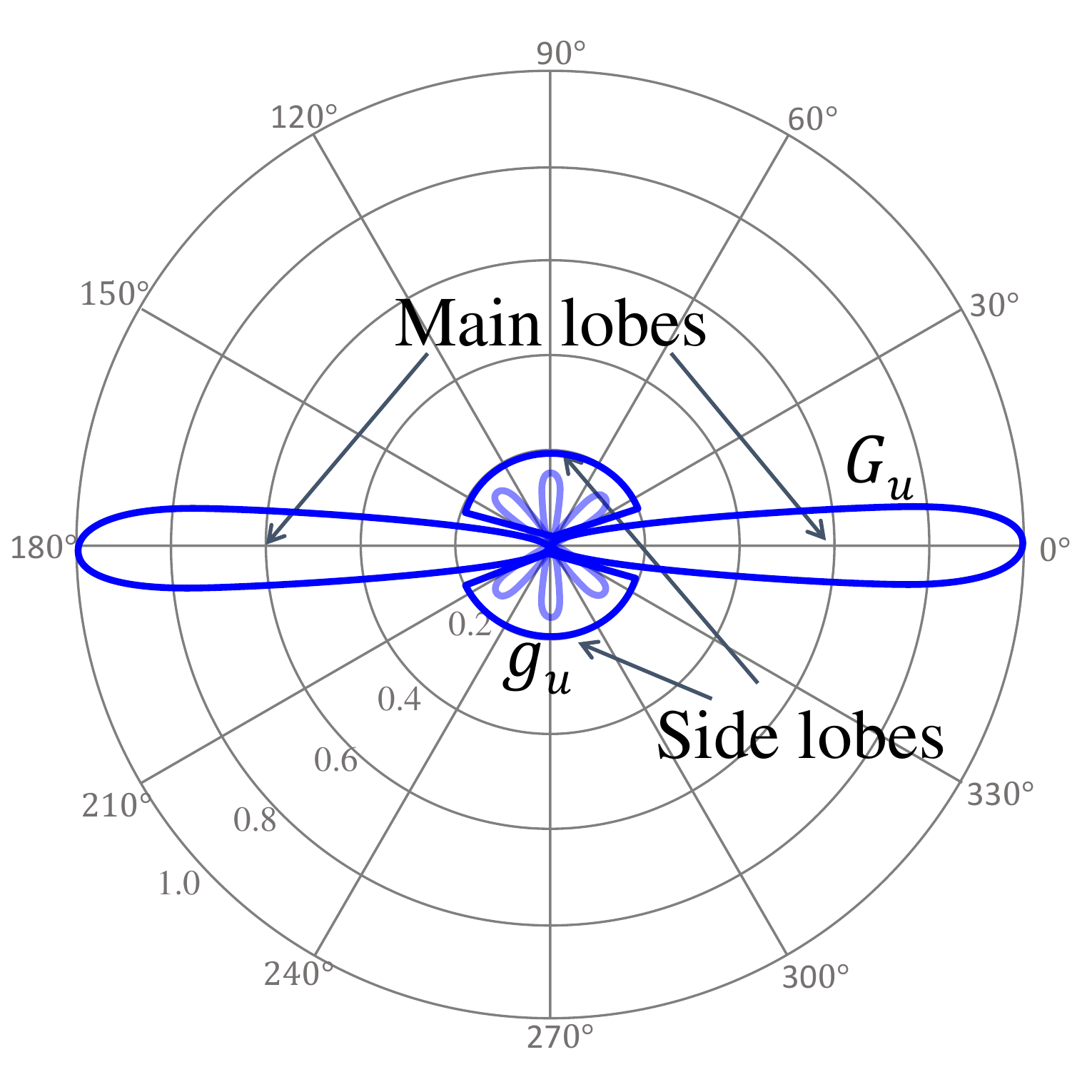}
	\caption{Illustration of the antenna radiation pattern for a vehicular node or an RSU.}
	\label{fig:beam_pattern_veh}
\end{minipage}
\quad 
\begin{minipage}{.38\textwidth}
	\centering
	\includegraphics[width=.7\textwidth]{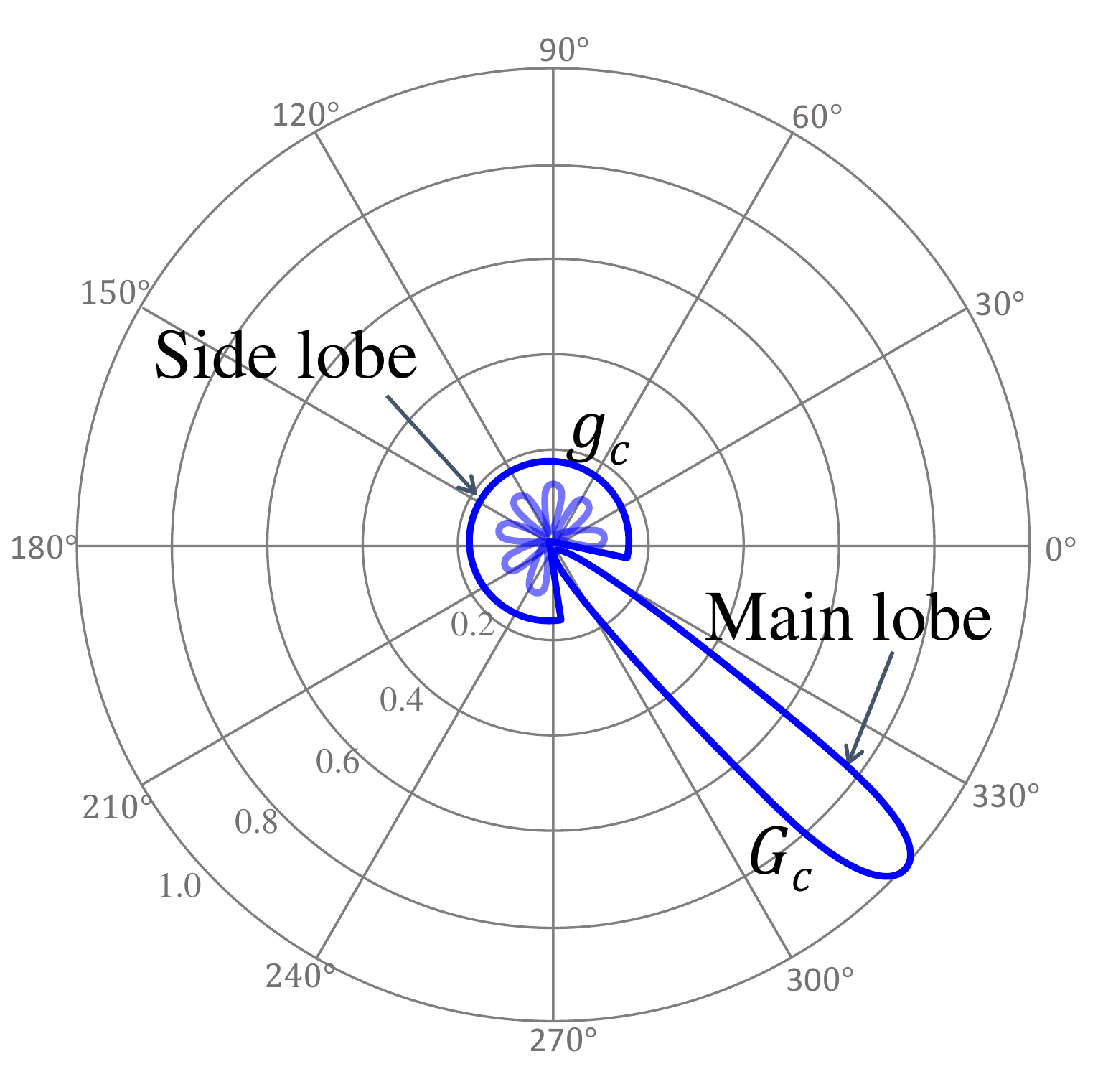}
	\caption{Illustration of the antenna radiation pattern for a cellular MBS.}
	\label{fig:beam_pattern_mbs}
\end{minipage}
\end{figure}

As the vehicular nodes and RSUs have identical transmit powers and beamforming patterns, we consider them as wireless nodes of a single tier in the coverage and rate analysis presented in this paper. Therefore, the vehicular nodes and RSUs will henceforth be collectively referred to as {\em tier 2 nodes} and the cellular MBSs will be referred to as {\em tier 1 nodes}. Thus, under Palm probability of the receiver point process, the locations of the tier 2 nodes form a Cox process $\Phi_2$ driven by the line process $\Phi_{l_0}$, where the locations of nodes on each line $L_j$ form a homogeneous 1D PPP $\Xi_{L_j}$ with density $\lambda_2 = \lambda_u + \lambda_v$. In order to be consistent with the notation, we denote the set of locations of tier 1 nodes by $\Phi_1$ with density $\lambda_1 = \lambda_c$. We also update the notation of transmit powers, main lobe and side lobe gains of tier 1 and tier 2 nodes to $\{P_1, G_1, g_1\}$ and $\{P_2, G_2, g_2\}$, respectively.

As the channel fading characteristics could vary significantly from rural to urban areas, we choose Nakagami-$m$ fading to mimic a wide range of fading environments. We denote the fading parameter for the link between the typical receiver and tier 1 nodes by $m_1$ and the channel fading gains by $H_1$. The tier 2 nodes on the typical line have a higher likelihood of a line-of-sight (LOS) link to the typical receiver than those nodes located on the other lines. So, in order to distinguish the severity of fading, we denote the fading parameters for the links to tier 2 nodes on the typical line and the tier 2 nodes on other lines by $m_{20}$ and $m_{21}$, and the corresponding fading gains by $H_{20}$ and $H_{21}$, respectively. The fading gains $H_i$, where $i \in\{1, 20, 21\}$ follow a Gamma distribution with probability density function (PDF) 
\begin{align}
f_{H_{i}} (h) = \frac{m_i^{m_i} h^{m_i-1}}{\Gamma(m_i)} \exp(-m_i h), \qquad m \in \{m_{1}, m_{20}, m_{21}\}.
\end{align}
In the interest of analytical tractability, we restrict the values of $m_{1}$, $m_{20}$ and $m_{21}$ to integers. 

We will now discuss the modeling of large scale fading effects. We assume a standard power-law path loss model with exponent $\alpha > 2$ for all the wireless links. We model the shadowing effects for the links from the typical receiver to the tier 1 nodes, tier 2 nodes on the typical line, and tier 2 nodes on other lines by the random variables $\calX_1$, $\calX_{20}$, and $\calX_{21}$, respectively. We assume that $\calX_i$, where $i \in \{1, 20, 21\}$, follows a log-normal distribution such that $10 \log_{10}\calX_i \sim  \mathcal{N}(\omega_i, \sigma_i^2)$ with mean $\omega_i$ and standard deviation $\sigma_i$ in dB. Thus, the received signal power at the typical receiver from a node located at $\nrmx$ is 
\begin{align}\label{eq:recd_power}
P_r (\nrmx) = \begin{dcases}
P_1 G_{\nrmx}  H_{1} \calX_{1} \|\nrmx \|^{-\alpha},  &\nrmx \in \Phi_1, \\
P_2 G_2  H_{20} \calX_{20} \|\nrmx \|^{-\alpha},  & \nrmx \in \Xi_{L_0}, \\
P_2 g_2  H_{21} \calX_{21} \|\nrmx \|^{-\alpha},  & \nrmx \in \Phi_2 \setminus \Xi_{L_0} ,\\
\end{dcases}
\end{align}
where $G_{\nrmx} \in \{ G_1 , g_1\}$ denotes the beamforming gain from the tier 1 node located at $\nrmx$, and $\| \cdot \|$ denotes the Euclidean norm. Note that we have assumed that the beamforming gain from all the tier 2 nodes that are not located on the typical line would be $g_2$. However, if the typical receiver is located \textit{exactly} at the intersection of two roads, then there would be two lines which contain nodes with a beamforming gain of $G_2$. This case can be handled by slightly modifying the equation for received power given in \eqref{eq:recd_power} and then following the same analytical procedure. However, since it is a zero measure event in our model, we do not have to consider it explicitly in this paper.




\subsection{Association Policy}
We assume that the typical receiver connects to the node which yields the highest average biased received power. As mentioned earlier, tier 1 nodes scan the space for the best beam alignment during the association process. 
Therefore, the candidate serving node from tier 1 is the node located at $\nrmx_1^* = \argmax_{\nrmx \in \Phi_1} P_1 G_1 \calX_{1} \| \nrmx\|^{-\alpha}$. A well-known procedure in the literature to handle the effect of shadowing is to express it as a random displacement of the location of the node w.r.t. the typical receiver \cite{Dhi_letter}. Thus, the average received power at the typical receiver from a tier 1 node can be alternately written as $P_r(\nrmy) = P_1 G_1 \| \nrmy \|^{-\alpha}$, where $\nrmy = \calX_1^{-\frac{1}{\alpha} } \nrmx$. By displacement theorem \cite{haenggi}, for a homogeneous PPP $\Psi \subset \nbbR^d$ with density $\lambda$, if each point $\nrmx_k \in \Psi$ is transformed to $\nrmy = \calX_k^{-\frac{1}{\alpha}} \nrmx_k$, then the transformed points also form a PPP with density $\nbbE[\calX^{-\frac{d}{\alpha}} ]\lambda$.
Thus, the candidate serving node from tier 1 is the node located at $\nrmy_1^* =  \argmax_{\nrmy \in \Phi_1^{(e)}} P_1 G_1  \| \nrmy\|^{-\alpha}$, where $\Phi_1^{(e)}$ is the equivalent 2D PPP with density $\lambda_1^{(e)} = \nbbE[\calX_1^{-\frac{2}{\alpha}} ]\lambda_1 = \exp\left(\frac{\omega_1\ln10 }{5 \alpha} + \frac{1}{2} \left(\frac{\sigma_1\ln 10 }{5 \alpha}\right)^2\right) \lambda_1$. Therefore, the candidate serving node from tier 1 is the closest node to the typical receiver from the point process $\Phi_1^{(e)}$.

  
Among the tier 2 nodes, the information obtained at the typical receiver from the nodes on the typical line is more relevant than the information from the nodes on the other lines. 
So, we consider that the typical receiver would connect to only those tier 2 nodes that lie on the typical line.
Therefore, the candidate tier 2 serving node is the node located at $\nrmx_2^* = \argmax_{\nrmx \in \Xi_{L_0}} P_2 G_2 \calX_{20} \|\nrmx \|^{-\alpha}$. As shown in case of tier 1 nodes, the impact of shadowing can be expressed as a random displacement of the location of the node at $\nrmx \in  \Xi_{L_0}$. Therefore, the candidate serving node from tier 2 is the node located at $\nrmy_2^* = \argmax_{\nrmy \in \Xi_{L_0}^{(e)}} P_2 G_2 \|\nrmy \|^{-\alpha}$, where $\Xi_{L_0}^{(e)}$ is the equivalent 1D PPP with density $\lambda_{2}^{(e)}=\nbbE[\calX_{20}^{-\frac{1}{\alpha}} ]\lambda_2$. Thus, the candidate serving node from tier 2 is the closest node to the typical receiver from the point process $\Xi_{L_0}^{(e)}$.

Therefore, the typical receiver either connects to its closest tier 1 node from $\Phi_1^{(e)}$ or the closest tier 2 located on the typical line $\Xi_{L_0}^{(e)}$. Further, we introduce selection bias factors $B_1$ and $B_2$ for the tier 1 and tier 2 nodes, respectively, in order to balance the load between them across the network. So, among the two candidate nodes from tier 1 and tier 2, the typical receiver connects to the node at $\nrmy^* = \argmax_{\nrmy \in \{\nrmy_j^* \}} B_j P_j G_j \|\nrmy\|^{-\alpha}$, where $j \in \{1,2\}$. 

For expositional simplicity, we assume that the system is interference limited and hence, the thermal noise is neglected in our analysis. The aggregate interference at the typical receiver is composed of the interference from the tier 1 nodes $I_1$, interference from the tier 2 nodes on the typical line $I_{20}$, and the interference from the tier 2 nodes on the other lines $I_{21}$. Thus, the SIR measured at the typical receiver is
\begin{align} 
&\sir = \frac{P_r(\nrmy^*)}{ I_1 + I_{20} + I_{21} } ,
\\\notag 
\text{where \hfill}& 
\\
\label{eq:i1}&I_1 = \sum\limits_{\scalebox{1}{${\mathclap{\substack{ \nrmx \in \Phi_1 \\  P_1 G_{\nrmx}  H_{1}\calX_{1}  \|\nrmx\|^{-\alpha} < P_r(\nrmy^*) } } }$}}  P_1 G_{\nrmx} H_{1}\calX_{1}  \|\nrmx\|^{-\alpha} =  \sum\limits_{\scalebox{1}{${\mathclap{\substack{ \nrmy \in \Phi_1^{(e)} \\  P_1 G_{\nrmy} H_{1}  \|\nrmy\|^{-\alpha} < P_r(\nrmy^*) } } }$}}  P_1 G_{\nrmy} H_{1}  \|\nrmy\|^{-\alpha}; G_{\nrmy} = \begin{dcases}
G_1 \text{ with prob. } q_c,\\
g_1 \text{ with prob. } 1-q_c.
\end{dcases}\\
\label{eq:i20}& I_{20} = \sum\limits_{\scalebox{1}{${\mathclap{\substack{ \nrmx \in \Xi_{L_0} \\  P_2 G_2 H_{20}\calX_{20}  \|\nrmx\|^{-\alpha} < P_r(\nrmy^*) } } }$}}  P_2 G_2 H_{20}\calX_{20}  \|\nrmx\|^{-\alpha} =  \sum\limits_{\scalebox{1}{${\mathclap{\substack{ \nrmy \in \Xi_{L_0}^{(e)} \\  P_2 G_2 H_{20}  \|\nrmy\|^{-\alpha} < P_r(\nrmy^*) } } }$}}  P_2 G_2 H_{20}  \|\nrmy\|^{-\alpha} ,
\\ 
\notag \text{and \hfill }&\\
\label{eq:i21}&I_{21} = \sum\limits_{ \mathclap{\substack{ \nrmx \in \Phi_2 \setminus \Xi_{L_0}}}} P_2 g_2 H_{21} \calX_{21} \| \nrmx\|^{-\alpha} = \sum\limits_{L_j \in \{\Phi_{l_0} \backslash L_0\} } \hspace{1em} \sum\limits_{\scalebox{1}{${ \mathclap{\substack{ \nrmx \in \Xi_{L_j} } } }$}} P_2 g_2 \calX_{21} H_{21} \|\nrmx\|^{-\alpha} .
\end{align} 

While the terms $I_1$ and $I_{20}$ were simplified by the application of displacement theorem, it is not possible for $I_{21}$. The characterization of $I_{21}$ is the key challenge that needs to be handled carefully in the computation of coverage probability.

\section{Coverage Analysis}
In this section, we will determine the SIR-based coverage probability of the typical receiver in the network. Recall that the typical receiver would connect to its closest tier 1 node from the 2D PPP $\Phi_1^{(e)}$ or the closest tier 2 node from the 1D PPP $\Xi_{L_0}^{(e)}$ and we denote these events by $\calE_1$ and $\calE_2$, respectively. As a result, the interference measured at the typical receiver is different in these two cases and hence we handle them separately in our analysis. We first determine the probability of occurrence of these events and then characterize the desired signal power by computing the cumulative distribution function (CDF) of the serving distance $R$ conditioned on these events. 

\subsection{Association Probability}
In this subsection, we derive the probability of the occurrence of the events $\calE_1$ and $\calE_2$. We denote the distances from the typical receiver to its closest node in $\Phi_1^{(e)}$ and $\Xi_{L_0}^{(e)}$ by $R_1$ and $R_2$, respectively. As is well known in the literature, the CDF and PDF of $R_1$ and $R_2$ are given by
\begin{align}\label{eq:cdfr1}
\textrm{CDF: } \quad  &F_{R_1} (r_1) = 1 - \exp \left(- \lambda_1^{(e)} \pi r_1^2\right),  \qquad F_{R_2} (r_2) = 1 - \exp \left(-\lambda_2^{(e)} 2 r_2\right),\\\label{eq:pdfr2}
\textrm{PDF: } \quad & f_{R_1} (r_1) = 2 \pi \lambda_1^{(e)} r_1 \exp \left(- \lambda_1^{(e)} \pi r_1^2\right), \qquad f_{R_2}(r_2) = 2 \lambda_2^{(e)} \exp \left(-\lambda_2^{(e)} 2 r_2\right).
\end{align}
Using these distributions, we will now derive the association probabilities of the typical receiver with tier 1 and tier 2 nodes in the following Lemma. As the shadowing terms are already handled, this derivation follows from the standard procedure but is given for completeness.  

\begin{lemma}\label{lem:pe1}
	The probability of occurrence of the events $\calE_1$ and $\calE_2$ are
	\begin{align}
	\P (\calE_1) &=   1- \lambda_2^{(e)} \sqrt{\frac{1}{\lambda_1^{(e)}  \zeta_{21}^{-\frac{2}{\alpha}}}} 
	\exp \left[ \frac{(\lambda_2^{(e)})^2}{\lambda_1^{(e)} \pi \zeta_{21}^{-\frac{2}{\alpha}}}\right] 
	\erfc \left(\frac{\lambda_2^{(e)}}{\sqrt{\lambda_1^{(e)} \pi \zeta_{21}^{-\frac{2}{\alpha}}}}\right),\\
	\P (\calE_2) &=  \lambda_2^{(e)} \sqrt{\frac{1}{\lambda_1^{(e)}  \zeta_{21}^{-\frac{2}{\alpha}}}} 
	\exp \left[ \frac{(\lambda_2^{(e)})^2}{\lambda_1^{(e)} \pi \zeta_{21}^{-\frac{2}{\alpha}}}\right] 
	\erfc \left(\frac{\lambda_2^{(e)}}{\sqrt{\lambda_1^{(e)} \pi\zeta_{21}^{-\frac{2}{\alpha}}}}\right),
	\end{align}
	where $\zeta_{21} = \frac{P_2 B_2 G_2}{P_1 B_1 G_1}$.
\end{lemma}\vspace{-.5em}
\begin{IEEEproof}
	See Appendix \ref{app:pe1}.
\end{IEEEproof}


\subsection{Serving Distance Distribution}
The next key step in computing the coverage probability is to characterize the desired signal power at the typical receiver which depends on the serving distance. Therefore, we will now derive the serving distance distribution conditioned on the events $\calE_1$ and $\calE_2$ in the following Lemmas.

\begin{lemma}\label{lem:cdfre1}
	Conditioned on the event $\calE_1$, the PDF of the serving distance is given by
	\begin{align}\notag
	f_R(r|\calE_1) &=   \frac{2 \pi \lambda_1^{(e)} r \exp\left(-\pi \lambda_1^{(e)} r^2 - 2 \zeta_{21}^{\frac{1}{\alpha}}\lambda_2^{(e)}r\right)}{\P(\calE_1)} 
	\end{align}
\end{lemma}
\begin{IEEEproof}
	The conditional CDF of $R$ can be obtained as
	\begin{align*}
	F_R(r | \calE_1) &= 1 - \frac{\P (R>r, \calE_1)}{\P (\calE_1)} \stackrel{(a)}{=} 1 - \frac{1}{\P(\calE_1)}\P \left(R_1 > r, R_1 < \left(\frac{P_2 B_2 G_2}{P_1 B_1 G_1}\right)^{-\frac{1}{\alpha}} R_2\right)\\
	&\stackrel{(b)}{=} 1 - \frac{1}{\P (\calE_1)} \int_{\zeta_{21}^{\frac{1}{\alpha}}r}^{\infty} \P \left(r < R_1 < \zeta_{21}^{-\frac{1}{\alpha}} r_2 | R_2 \right) f_{R_2}(r_2) {\rm d} r_2\\
	&= 1 - \frac{1}{\P(\calE_1)} \int_{\zeta_{21}^{\frac{1}{\alpha}}r}^{\infty} \left[F_{R_1}\left(\zeta_{21}^{-\frac{1}{\alpha}}r_2\right) - F_{R_1}(r)\right]f_{R_2}(r_2) {\rm d} r_2,
	\end{align*}
	where (a) follows form the condition for the occurrence of the event $\calE_1$, (b) follows from substituting $\zeta_{21} = \frac{P_2 B_2 G_2}{P_1 B_1 G_1}$ and conditioning on $R_2$, and the final expression for CDF can be obtained by substituting the expressions for $F_{R_1}(\cdot)$ and $f_{R_2}(\cdot)$ in the last step and simplifying the resulting integral. The PDF of the serving distance conditioned on $\calE_1$ can then be obtained by taking the derivative of $F_R(r|\calE_1)$ w.r.t. $r$.
\end{IEEEproof}\vspace{-.5em}
\begin{lemma}\label{lem:cdfre2}
	Conditioned on the event $\calE_2$, the PDF of the serving distance is given by
	\begin{align}\notag
	 f_R(r|\calE_2) &=  \frac{2 \lambda_2^{(e)}   \exp \left(-2 \lambda_2^{(e)} r - \pi \lambda_1^{(e)}\zeta_{21}^{-\frac{2}{\alpha}} r^2 \right) }{\P \left(\calE_2\right)} ,
	\end{align}
	where $\zeta_{21} = \frac{P_2 B_2 G_2 }{P_1 B_1 G_1}$. 
\end{lemma}
\begin{IEEEproof}
	The proof follows along the same lines as that of Lemma \ref{lem:cdfre1} and hence skipped.
\end{IEEEproof}

\subsection{Approximation of the Cox Process Model}\label{sec:approx}

In order to determine the coverage probability, we need to compute the conditional Laplace transform of the interference power distribution. Recall that the aggregate interference at the typical receiver can be decomposed into three independent components $I_1$, $I_{20}$, and $I_{21}$, caused by tier 1 nodes, tier 2 nodes on the typical line and the tier 2 nodes on other lines, respectively. The key technical challenge in the computation of Laplace transform of interference stems from the inclusion of shadowing effects. As explained in section \ref{sec:systemmodel}, a well-known approach in the literature is to interpret the effect of shadowing as a random displacement of the location of the nodes and then compute the received power at the typical receiver from the nodes whose locations are given by the equivalent point process. 
So, we can apply this technique to the 2D PPP of tier 1 nodes $\Phi_1$ and 1D PPP of tier 2 nodes on the typical line $\Xi_{L_0}$ and rewrite the interference powers $I_1$ and $I_{20}$ in terms of the equivalent points processes $\Phi_1^{(e)}$ and $\Xi_{L_0}^{(e)}$, respectively, as given in \eqref{eq:i1} and \eqref{eq:i20}.
However, upon applying this technique to the Cox process of tier 2 nodes excluding the nodes on the typical line $\Phi_2' = \Phi_2 \setminus \Xi_{L_0}$, the collinearity of the locations of the nodes is disrupted in the resulting point process $\Phi_{2}'^{(e)} = \{ \nrmy: \nrmy = \calX_{21}^{-\frac{1}{\alpha}} \nrmx, \nrmx \in \Phi_2' \}$ due to the random displacement of each point on a line. 
Thus, in order to exactly characterize the interference, it is necessary to find the exact distribution of points in $\Phi_2'^{(e)}$, which is quite hard. This is the key motivation behind proposing a tractable yet accurate approximation for the interference analysis of this component.

From Theorem \ref{thm:convergencetoPPP}, we already know that the Cox process asymptotically converges to a 2D PPP. Further, as mentioned earlier, the effect of shadowing interpreted as random displacement of points also disturbs the collinear structure of points on all the lines except the typical line. Motivated by these two facts, we make the following assumption which enables a tractable interference analysis. \vspace{-.5em}
\begin{assumption}\label{assum:ppp}
	We assume that the spatial distribution of nodes in $\Phi_2'^{(e)}$ follows a 2D PPP with density $\nbbE[\calX_2^{-\frac{2}{\alpha}}]\pi \lambda_l \lambda_2$. This density follows from the result given in Theorem \ref{thm:convergencetoPPP} combined with the interpretation of shadowing as random displacement of nodes.
\end{assumption}

Under Assumption \ref{assum:ppp}, the approximate spatial model for the tier 2 nodes with the inclusion of shadowing effects is  $\tilde{\Phi}_2^{(e)} = \Xi_{L_0}^{(e)} \cup \Phi_2^{(a)}$, where $\Phi_{2}^{(a)}$ is a 2D PPP with density $\lambda_2^{(a) }= \nbbE[\calX_2^{-\frac{2}{\alpha}}]\pi \lambda_l \lambda_2$. {In other words}, the proposed approximation for the tier 2 nodes is the superposition of the 1D PPP $\Xi_{L_0}^{(e)}$ with density $\lambda_2^{(e)}$ on the typical line and the 2D PPP $\Phi_2^{(a)}$ with density $\lambda_{2}^{(a)}$. \vspace{-.75em}
\begin{remark}
It must be noted that the $K$-functions of $\{\Xi_{L_0}^{(e)} \cup \Phi_{2}'^{(e)}\}$ and  $\tilde{\Phi}_2^{(e)}$ are identical and are given by $K(r) = \frac{2r}{\pi \lambda_l} + \pi r^2$, where the first term corresponds to the points located on the typical line and the second term corresponds to the rest of the points. This is further proof that this is a reasonable approximation. More numerical evidence will be provided in Section \ref{sec:num}. 
\end{remark}\vspace{-.75em}
Now that we have clearly established the approximate spatial model for the tier 2 nodes, we proceed with the coverage analysis by computing the Laplace transform of interference in the next subsection.

\subsection{Laplace Transform of Interference Power Distribution}
In this subsection, we will compute the Laplace transform of the distribution of interference power measured at the typical receiver conditioned on the serving distance $R$ and the events $\calE_1$ and $\calE_2$. First, let us consider $\calE_1$. In case of $\calE_1$, as the typical receiver connects to its closest tier 1 node at a distance $R$, there can not be any tier 1 node closer than $R$. Also, the average received power from any tier 2 nodes on the typical line can not exceed the received power from the serving node. This means that there can not be any tier 2 node on the typical line whose distance to the typical receiver is lesser than $\zeta_{21}^{\frac{1}{\alpha}} R$. We have similar restrictions on the spatial distribution of nodes in case of $\calE_2$. Incorporating these conditions, we determine the Laplace transform of interference conditioned on $\calE_1$ and $\calE_2$ in the following Lemmas.
\vspace{-.5em}
\begin{lemma}\label{lem:lIe1}
	Conditioned on the event $\calE_1$ and the serving distance $R$, the Laplace transform of the interference at the typical receiver is 
	\begin{align}\notag 
	\calL_I&(s|R, \calE_1) = \exp \Bigg[ \scalebox{2.0}[1.0]{-}2 \pi q_c \lambda_1^{(e)} \int_r^{\infty} \mspace{-6mu} \Bigg(1 \scalebox{2.0}[1.0]{-} \left(1 + \frac{s P_1 G_1 y^{-\alpha}}{m_1}\right)^{-m_1} \mspace{-6mu} \Bigg)y {\rm d} y  \scalebox{2.0}[1.0]{-} 2 \pi (1- q_c) \lambda_1^{(e)} \\ \notag
	&\hspace{3em} \times \int_r^{\infty} \mspace{-6mu} \Bigg(1 \scalebox{2.0}[1.0]{-} \Bigg(1 + \frac{s P_1 g_1 y^{-\alpha}}{m_1}\Bigg)^{-m_1} \mspace{-6mu}\Bigg) y {\rm d}y \scalebox{2.0}[1.0]{-} 2  \lambda_2^{(e)} \int_{\zeta_{21}^{\frac{1}{\alpha}} r}^{\infty} \Bigg(1 \scalebox{2.0}[1.0]{-} \bigg(1 + \frac{s P_2 G_2 y^{-\alpha }}{m_{20}}\bigg)^{-m_{20}} \Bigg) {\rm d} y\\
	& \hspace{3em} \scalebox{2.0}[1.0]{-} 2 \pi \lambda_2^{(a)} \int_{0}^{\infty} \left(1 \scalebox{2.0}[1.0]{-} \bigg(1 + \frac{s P_2 g_2 y^{-\alpha }}{m_{21}}\bigg)^{-m_{21}} \right) y {\rm d} y \Bigg].
	\end{align}
\end{lemma}
\begin{IEEEproof}
See Appendix \ref{app:lIe1}.
\end{IEEEproof}

\begin{lemma}\label{lem:lIe2}
Conditioned on the event $\calE_2$ and the serving distance $R$, the Laplace transform of the interference at the typical receiver is 
\begin{align}\notag
\calL_I&(s|R, \calE_2) =  \exp \Bigg[ \scalebox{2.0}[1.0]{-}2 \pi q_c \lambda_1^{(e)} \int_{\zeta_{21}^{-\frac{1}{\alpha}} r}^{\infty} \mspace{-6mu} \Bigg(1 \scalebox{2.0}[1.0]{-} \left(1 + \frac{s P_1 G_1 y^{-\alpha}}{m_1}\right)^{-m_1} \mspace{-6mu} \Bigg)y {\rm d} y  \scalebox{2.0}[1.0]{-} 2 \pi (1- q_c) \lambda_1^{(e)} \\ \notag
&\hspace{3em} \times \int_{\zeta_{21}^{-\frac{1}{\alpha}} r}^{\infty} \mspace{-6mu} \Bigg(1 \scalebox{2.0}[1.0]{-} \Bigg(1 + \frac{s P_1 g_1 y^{-\alpha}}{m_1}\Bigg)^{-m_1} \mspace{-6mu}\Bigg) y {\rm d}y \scalebox{2.0}[1.0]{-} 2  \lambda_2^{(e)} \int_{r}^{\infty} \Bigg(1 \scalebox{2.0}[1.0]{-} \bigg(1 + \frac{s P_2 G_2 y^{-\alpha }}{m_{20}}\bigg)^{-m_{20}} \Bigg) {\rm d} y\\
& \hspace{3em} \scalebox{2.0}[1.0]{-} 2 \pi \lambda_2^{(a)} \int_{0}^{\infty} \left(1 \scalebox{2.0}[1.0]{-} \bigg(1 + \frac{s P_2 g_2 y^{-\alpha }}{m_{21}}\bigg)^{-m_{21}} \right) y {\rm d} y \Bigg].
\end{align}
\end{lemma}
\begin{IEEEproof}
	The proof follows along the same lines as that of Lemma \ref{lem:lIe1} and hence skipped.
\end{IEEEproof}

\subsection{Coverage Probability}
The coverage probability is formally defined as the probability with which the SIR measured at the typical receiver exceeds a predetermined threshold $\beta$. Using the conditional distributions of serving distance and the conditional Laplace transform of interference power distribution, we derive the coverage probability in the following theorem.
\begin{thm}\label{thm:pc}
	The coverage probability of the typical receiver is
\begin{align}\notag 
\pc &=\P (\calE_1) \sum_{k=0}^{m_1-1} \int_0^{\infty}  \frac{(\scalebox{2.0}[1.0]{-} m_1 \T)^k}{ r^{- k \alpha} (P_1 G_1)^k k!} \bigg[ \frac{\partial^k}{\partial s^k} \calL_I(s| R, \calE_1) \bigg]_{s=\frac{m_1 \T r^{\alpha}}{P_1 G_1}} f_R(r|\calE_1) {\rm d}r\\  
&\hspace{4em}+\P (\calE_2) \sum_{k=0}^{m_{20}-1} \int_0^{\infty}  \frac{(\scalebox{2.0}[1.0]{-} m_{20} \T)^k}{ r^{- k \alpha} (P_2 G_2)^k k!} \bigg[ \frac{\partial^k}{\partial s^k} \calL_I(s| R, \calE_2) \bigg]_{s=\frac{m_{20} \T r^{\alpha}}{P_2 G_2}} f_R(r|\calE_2) {\rm d}r,
\end{align}
where $\P(\calE_1)$ and $\P(\calE_2)$ are given in Lemma \ref{lem:pe1}, $\calL_{I}(s|R,\calE_1)$ and $\calL_{I}(s|R,\calE_2)$ are given in Lemmas \ref{lem:lIe1} and \ref{lem:lIe2}, and $f_R(r|\calE_1)$ and $f_R(r|\calE_2)$ are given in Lemmas \ref{lem:cdfre1} and \ref{lem:cdfre2}. 
\end{thm}
\begin{IEEEproof}
	The proof follows along the same lines as that of Theorem 1 in \cite{vishnuJ2} and is hence skipped.
\end{IEEEproof}

\section{Downlink Rate Coverage}

 In this section, we will compute the downlink rate coverage of the typical receiver by characterizing the load on the serving tier 1 and tier 2 nodes. {This characterization of load is one of the key contributions of this paper and is useful in studying several other metrics such as latency and packet reception rate \cite{3gpp}.} Assuming that the bandwidth resource is equally shared by all the users that are connected to the serving node, the downlink rate achievable at the typical receiver can be computed using Shannon's theorem as 
 \begin{align}\label{eq:ratecov1}
 \calR = \frac{W}{J} \log_2\left( 1 + \sir \right)  \textrm{    bits/sec},
 \end{align}
 where $W$ is the available bandwidth and $J$ is the load on the serving node of the typical receiver. We will refer to the serving node of the typical receiver as the \textit{tagged node} and the corresponding cell as the \textit{tagged cell}. We denote the load on the tagged tier 1 and tier 2 nodes by $J_1$ and $J_2$, respectively. As we have already characterized the $\sir$ at the typical receiver in the previous section, we will now focus on the distribution of $J_1$ and $J_2$ in the following subsections.

 
\subsection{Load on the Tagged Tier 2 Node}

As the load on a node is proportional to the size of its cell, we need to characterize the size of the tagged tier 2 cell. First, we will focus on the cell of a \textit{typical} tier 2 node, which is an arbitrarily chosen tier 2 node. We assume that the typical tier 2 node is located at ${\nrmx_{\rm typ}}$ on a line $L$. As the tier 2 nodes serve only the users located on their own lines, the typical cell is a line segment on the line $L$ and is denoted by $\calZ_{\rm typ} = \{ \nrmx: \nrmx \in L,  P_2 G_2 B_2 \|\nrmx- \nrmx_{\rm typ}\|^{-\alpha} \geq P_\nrmy G_\nrmy B_\nrmy \| \nrmx - \nrmy \|^{-\alpha}, \forall \nrmy \in \Phi_1^{(e)} \cup \Xi_{L}^{(e)} \}$, where $P_\nrmy$, $G_\nrmy$, and $B_\nrmy$ correspond to the transmit power, beamforming gain, and bias of the node located at $\nrmy$, respectively. We begin our analysis by computing the distribution of the length of the typical tier 2 cell, which is denoted by $Z_{\rm typ} = \nu_1 (\calZ_{\rm typ})$.

\begin{lemma}\label{lem:pdfztyp}
	The PDF of the length of a typical tier 2 cell is 
	\begin{align}\notag
	f_{Z_{\rm typ}}(z) &\approx \int_{\frac{k+1}{2k}z}^{z} f_{Z_1,1}(z - z_0) f_{Z_0}(z_0) {\rm d}z_0  + \int_{\frac{k-1}{2k}z}^{\frac{k+1}{2k}z} f_{Z_1,2}(z - z_0) f_{Z_0}(z_0) {\rm d}z_0  \\
	& \hspace{18em} + \int_0^{\frac{k-1}{2k}z} f_{Z_1,3}(z - z_0) f_{Z_0}(z_0) {\rm d}z_0,
	\end{align}
	where 
	\begin{align*}
	&f_{Z_1,1}(z_1) = 2 \lambda_2^{(e)}  \exp(\scalebox{2.0}[1.0]{-} 2 \lambda_2^{(e)} z_1),   \hspace{4em} 0<z_1 < \frac{k-1}{k+1}z_0, 
	\\
	&f_{Z_1,2}(z_1) = (2 \lambda_2^{(e)} + \lambda_1^{(e)} \frac{\partial \gamma_{2,2}(z_0, z_1) }{\partial z_1}  )\exp\bigg[ \scalebox{2.0}[1.0]{-} 2 \lambda_2^{(e)} z_1 \scalebox{2.0}[1.0]{-}  \lambda_1^{(e)} \gamma_{2,2}(z_0, z_1) \bigg] ,  \ \frac{k-1}{k+1}z_0 < z_1 < \frac{k+1}{k-1}z_0,
	\\
	&f_{Z_1,3}(z_1) = (2  \lambda_2^{(e)} +  2\lambda_1^{(e)} \pi k^2 z_1 )\exp \bigg[ \scalebox{2.0}[1.0]{-}  2  \lambda_2^{(e)} z_1 \scalebox{2.0}[1.0]{-}  \lambda_1^{(e)} \pi k^2 z_1^2 + \lambda_1^{(e)} \pi k^2 z_0^2 \bigg], \ \frac{k+1}{k-1}z_0 < z_1 < \infty,\\
	&f_{Z_0}(z_0) = \left(2 \lambda_2^{(e)} + 2 \lambda_1^{(e)} \pi \zeta_{21}^{-\frac{2}{\alpha}} z_0 \right) \exp\left[- 2 \lambda_2^{(e)} z_0 - \lambda_1^{(e)} \pi\zeta_{21}^{-\frac{2}{\alpha}} z_0^2\right],\\
	 &\gamma_{2,2}(z_0, z_1) = \pi k^2 z_1^2 - (k z_0)^2 (\theta -\frac{1}{2} \sin(2\theta) ) - (k z_1)^2 (\phi -\frac{1}{2} \sin(2\phi)),\\
	 &\theta= \arccos \Bigg(\frac{(z_0 + z_1)^2 + (k z_0)^2 - (k z_1)^2}{2 k z_0 ( z_0 + z_1)}\Bigg), \  \phi = \arccos \Bigg( \frac{(z_0 + z_1)^2 - (k z_0)^2 + (k z_1)^2}{2 k z_1 ( z_0 + z_1)}\Bigg), \\
	&\text{and} \  k = \zeta_{21}^{-\frac{1}{\alpha}}.
	\end{align*}
\end{lemma}
 \begin{IEEEproof}
 See Appendix \ref{app:pdfztyp}.
 \end{IEEEproof}

In order to compute the load on the tagged tier 2 node located at $\nrmy_2^*$, we now need to determine the length of the tagged tier 2 cell. Recall that the tagged tier 2 cell is the cell that contains the typical receiver located at the origin and is denoted by $\calZ_{\rm tag} = \{ \nrmx: \nrmx \in L_0,  P_2 G_2 B_2 \|\nrmx- \nrmy_2^*\|^{-\alpha} \geq P_\nrmy G_\nrmy B_\nrmy \| \nrmx - \nrmy \|^{-\alpha}, \forall \nrmy \in \Phi_1^{(e)} \cup \Xi_{L_0}^{(e)} \}$. Hence, the length of the tagged tier 2 cell is larger on average than that of a typical tier 2 cell. Thus, the PDF of the length of the tagged tier 2 cell $Z_{\rm tag}$ can be computed from the PDF of the length of a typical tier 2 cell, as given in the following Lemma.
\begin{lemma}\label{lem:pdftag}
	The PDF of the length of the tagged tier 2 cell is 
	\begin{align}
	    f_{Z_{\rm tag}}(w) = \frac{w}{\int_0^{\infty} w f_{Z_{\rm typ}}(w) {\rm d}w} f_{Z_{\rm typ}}(w),
	\end{align}
	where $f_{Z_{\rm typ}}(\cdot)$ is given in Lemma \ref{lem:pdfztyp}.
\end{lemma} 
\begin{IEEEproof}
	The proof simply follows from the length-biased sampling \cite{patil, offlsshd}.
\end{IEEEproof} 
 
Having determined the PDF of the tagged tier 2 cell length, we now compute the distribution of load on the tagged tier 2 node in the following Lemma.
 \begin{lemma}\label{lem:pmfload}
 	The probability mass function (PMF) of the load on the tagged tier 2 node is 
 	\begin{align}
 	\P ( J_2 = j_2 + 1| \calE_2 ) =  \int_0^{\infty} \frac{\exp(- \lambda_r w) (\lambda_r w)^{j_2}}{j_2!} f_{Z_{\rm tag}}(w) {\rm d}w, \qquad j_2 = 0, 1, 2, \dots
 	\end{align}
 \end{lemma}
 \begin{IEEEproof}
 	The proof follows from the Poisson distribution of receiving vehicular user nodes on the typical line. 
 \end{IEEEproof}
 
\subsection{Load on Tagged Tier 1 Node}
\begin{figure}
	\centering
	\includegraphics[width=.5\textwidth]{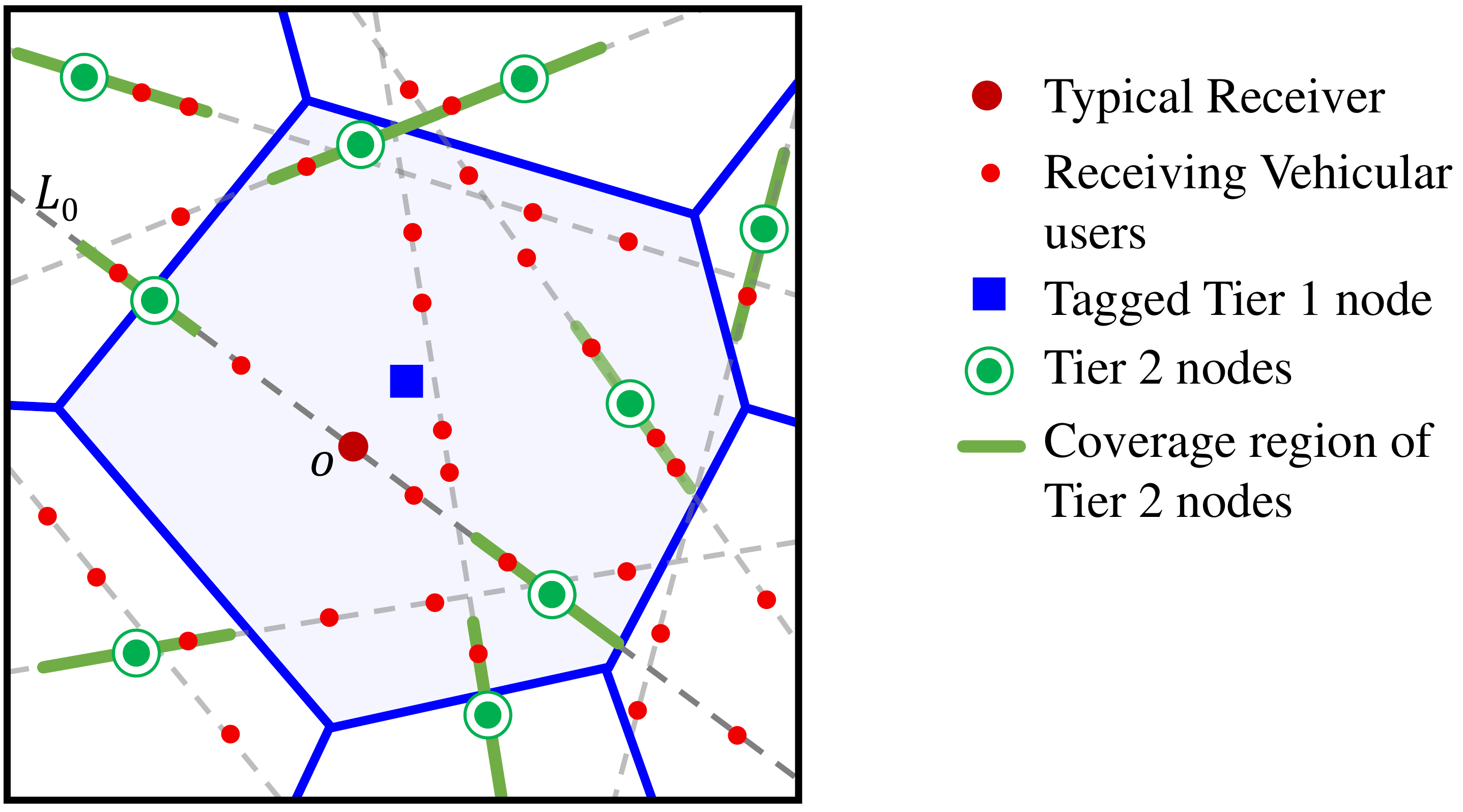}
	\caption{Illustration of the coverage region of a tagged tier 1 node.}
	\label{fig:cov_zones}
\end{figure}
In contrast to the tier 2 nodes with 1D coverage regions, the tier 1 nodes have 2D coverage regions which partition $\nbbR^2$ into multiple cells, as shown in Fig. \ref{fig:cov_zones}. As all the tier 1 nodes have the same transmit power and beamforming gains during the association process, their coverage regions are simply given by the Voronoi cells constructed from the 2D PPP $\Phi_1^{(e)}$. Thus, the tagged tier 1 cell is the Voronoi cell containing the origin and is denoted by $\calV_{\rm tag} = \{ \nrmx: \|\nrmx-\nrmy_1^*\| \leq  \|\nrmx-\nrmy\| \forall \nrmy \in \Phi_1^{(e)}  \}$. Although the vehicular users located on the line segments inside $\calV_{\rm tag}$ receive the highest average biased power among the tier 1 nodes from the node at $\nrmy_1^*$, some of the vehicular users inside the cell are served by tier 2 nodes located on these line segments, as illustrated in Fig. \ref{fig:cov_zones}. Thus, a few segments of the lines inside the Voronoi cell are not part of the tagged cell and hence, the load on the tagged tier 1 node is the number of vehicular users located on the line segments inside $\calV_{\rm tag}$ which are not covered by any tier 2 node. As the load is proportional to the length of these line segments, we need to compute the exact distribution of the total length of line segments that are not covered by tier 2 nodes inside $\calV_{\rm tag}$, which is intractable. Hence, we propose to compute the rate coverage using the mean load on the tagged node, which will be formally explained in the next subsection. {So, we will now compute the mean load on the tagged tier 1 node in this network, which has not been studied in the literature. This result will be derived using the results discussed in the previous subsection.} 
 
\begin{lemma}\label{lem:barJ1}
	The mean load on a tagged tier 1 node is  
	\begin{align}
	\bar{J_1} = 1 + \lambda_r \left( \frac{1.28 \pi \lambda_l  }{\lambda_1^{(e)}}+ \frac{3.216}{\pi \sqrt{\lambda_1^{(e)}}}\right)  \left( 1 - \lambda_2^{(e)} \int_0^{\infty} z f_{Z_{\rm typ}} (z) {\rm d}z \right).
	\end{align}
\end{lemma}
 \begin{IEEEproof}
 	See Appendix \ref{app:meanload}.
%
  \end{IEEEproof}

\subsection{Rate Coverage}

Using the results derived thus far, we will now compute the rate coverage of the typical receiver which is formally defined as the probability with which the data rate achievable at the typical receiver exceeds a predefined target rate $T$. Before we proceed to the computation of overall rate coverage, we will first focus on the case where the typical receiver is connected to a tier 1 node. As discussed in the previous subsection, it is quite hard to exactly characterize the load on the tagged tier 1 node. So, we make the following assumption which enables the computation of the rate coverage without much loss in the accuracy of our final results.
\begin{assumption}\label{assum:meanload}
	We assume that all the tier 1 nodes are operating at the mean load.
\end{assumption}
Under this assumption, the overall rate coverage of the typical receiver is given in the following theorem.
\begin{thm} 
The rate coverage of the typical receiver is
\begin{align}
\rc = \P \left(\sir > 2^{\frac{T \bar{J_1}}{W}} -1 | \calE_1 \right) \P(\calE_1) +  \P(\calE_2)\sum_{j_2 = 1}^{\infty} \P \left(\sir > 2^{\frac{T j_2}{W}} \scalebox{2.0}[1.0]{-} 1 | \calE_2, J_2 \right) \P (J_2 = j_2 | \calE_2) . 
\end{align}
\end{thm}
\begin{IEEEproof}The rate coverage can be computed as
\begin{align}\notag 
\rc &= \P ( \calR > T) = \sum_{i \in \{1, 2\} } \P ( \calR > T | \calE_i) \P (\calE_i) = \sum_{i \in \{1, 2\} } \nbbE_{J_i} \bigg[ \P ( \calR > T | \calE_i, J_i) \bigg] \P (\calE_i) \\ \label{eq:ratecov}
&= \sum_{i \in \{1, 2\} } \sum_{j_i = 1}^{\infty} \P \left(\sir > 2^{\frac{T j_i}{W}} -1 | \calE_i, J_i \right) \P (J_i = j_i | \calE_i) \P(\calE_i).
\end{align}
Although the $\sir$ measured at the typical receiver and the load on the tagged node are correlated, we assume that they are independent in the interest of analytical tractability and proceed with our analysis. 
Thus, the rate coverage is given by
\begin{align}\notag
\rc &= \sum_{i \in \{1, 2\} } \sum_{j_i = 1}^{\infty} \P \left(\sir > 2^{\frac{T j_i}{W}} -1 | \calE_i\right) \P (J_i = j_i | \calE_i) \P(\calE_i)\\
&\stackrel{(a)}{=} \P \left(\sir > 2^{\frac{T \bar{J_1}}{W}} -1 | \calE_1 \right) \P(\calE_1) +  \P(\calE_2)\sum_{j_2 = 1}^{\infty} \P \left(\sir > 2^{\frac{T j_2}{W}} \scalebox{2.0}[1.0]{-} 1 | \calE_2, J_2 \right) \P (J_2 = j_2 | \calE_2), 
\end{align}
where (a) follows from Assumption \ref{assum:meanload}. This completes the proof.
\end{IEEEproof}

\section{Numerical Results and Discussion} \label{sec:num}

In this section, we will present the numerical results for the coverage and rate analysis of the vehicular network. We will verify the accuracy of our analytical results by comparing them with the empirical results obtained from Monte-Carlo simulations. We will also discuss the effect of various parameters such as selection bias factors, and densities of nodes on the performance of the network. We will then provide a few design insights that could aid in improving the performance of the network without deploying additional infrastructure.

\subsection{Coverage Probability}
We first simulate the network model described in section \ref{sec:systemmodel} in MATLAB to compute the empirical distribution of coverage probability and rate coverage. For all the numerical results presented in this section, we assume that $\lambda_r = 15$ nodes/km, $\alpha = 4$, $W = 10$ MHz, $m_1 = m_{20} = m_{21} = 1$ and $\omega_1= \omega_{20} = \omega_{21} = 0$ dB. We assume that the ratio of main lobe to side lobe gains for both tier 1 and tier 2 nodes is 20 dB. We also assume that the beamwidth of the main lobe of a tier 1 node is $0.1 \pi$ and its orientation is uniformly distributed in $[0, 2\pi)$. Thus, the probability $q_c$ that the main lobe of an interfering tier 1 node is aligned towards the typical receiver is 0.05. We verify our analytical results for coverage probability by numerically comparing them with the empirical results in Fig. \ref{fig:pc_1}. As expected, the coverage probability evaluated using the analytical expressions in Theorem \ref{thm:pc} matches closely with the results obtained from simulations. This shows that the approximation of the Cox process model presented in Section \ref{sec:approx} is remarkably accurate for the coverage analysis. Therefore, the proposed approximation could also enable other similar analyses that may not otherwise be possible due to the nature of the Cox process.

{\em Impact of selection bias.} We plot the coverage probability of the typical receiver as a function of selection bias of tier 2 nodes $B_2$ in Fig. \ref{fig:pc_2}. Although the coverage probability improves initially with an increase in $B_2$, it begins to degrade beyond a certain value of $B_2$ as the interference power from tier 1 nodes dominates the desired signal power. Hence, there exists an optimal value of $B_2$ that maximizes the coverage probability of the typical receiver for given tier 1 and tier 2 node densities, as shown in Fig. \ref{fig:pc_2}.

{\em Impact of node densities.} From Fig. \ref{fig:pc_2}, we can also observe that the coverage probability of the typical receiver decreases as the density of tier 2 nodes increases when there is no selection bias ($B_1=B_2$ = 0 dB). Although a denser distribution of tier 2 nodes would increase both the desired signal power and the interference power when the typical receiver is connected to a tier 2 node, it would increase only the interference power when the typical receiver is associated with a tier 1 node. Hence, the overall coverage probability decreases with increasing tier 2 node density. However, this trend changes depending on the selection bias as we observe that the curves corresponding to different densities cross over beyond a certain value of $B_2$, as shown in Fig. \ref{fig:pc_2}. Similarly, from Fig. \ref{fig:pc_3}, we observe that the coverage probability increases with an increase in the density of tier 1 nodes when there is no bias and the trend gradually changes as the selection bias $B_1$ increases. 
\begin{figure}
\begin{minipage}{.45\textwidth}
	\centering
	\includegraphics[width=\textwidth]{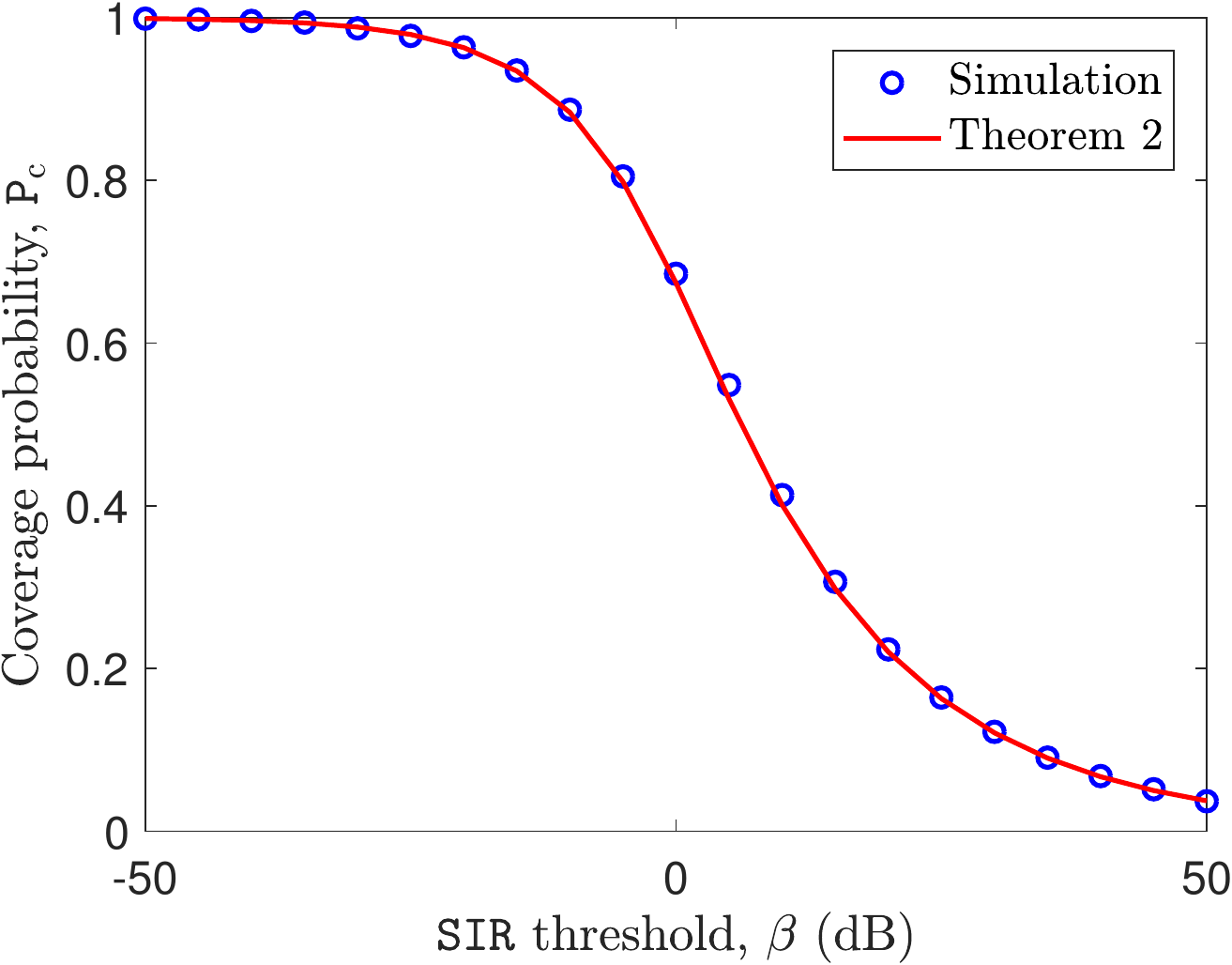}
	\caption{Coverage probability of the typical receiver as a function of $\sir$ threshold ($\mu_l$ = 10 km$^{-1}$, $\lambda_1 = 0.5$ nodes/km$^2$, $\lambda_2 = 4$ nodes/km, $B_1 = B_2 = 0 $ dB, $P_1 = 40$ dBm, $P_2 = 23$ dBm and $[\sigma_1\ \sigma_{20}\ \sigma_{21} ] = [ 4\ 2\ 4]$ dB).}
	\label{fig:pc_1}
\end{minipage}
\hfill
\begin{minipage}{.45\textwidth}
	\centering
	\includegraphics[width=\textwidth]{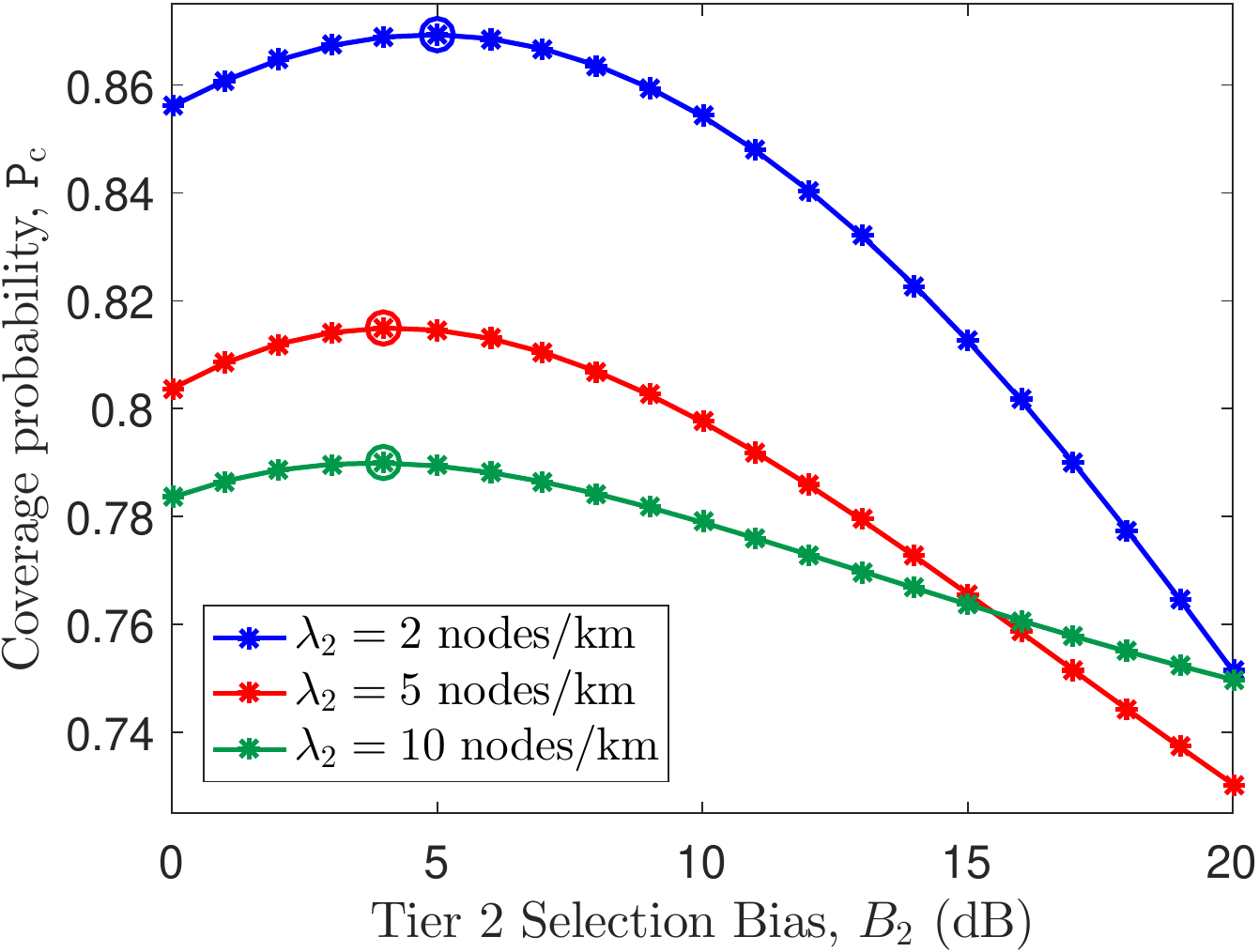}
	\caption{Coverage probability of the typical receiver as a function of selection bias $B_2$ ($\mu_l$ = 5 km$^{-1}$, $\lambda_1 = 2$ nodes/km$^2$, $B_1 = 0 $ dB, $P_1 = 43$ dBm, $P_2 = 23$ dBm, $[\sigma_1\ \sigma_{20}\ \sigma_{21} ] = [ 4\ 2\ 4]$ dB, and $\beta = 0$ dB).}
	\label{fig:pc_2}
\end{minipage}
\end{figure}

\subsection{Rate Coverage}
As the distribution of load is a key component in the computation of the downlink rate, we first compare the CDF of the load on the tagged tier 2 node evaluated using the expression given in Lemma \ref{lem:pmfload} with the results obtained from simulations. We observe that the theoretical results match closely with the simulations as shown in Fig. \ref{fig:loadcomp}. This also shows that the PDF of the length of the tagged tier 2 cell given in Lemma \ref{lem:pdftag} is quite accurate. We observe that the approximate load distribution, combined with the exact association probabilities and accurate coverage probability expressions, yields a tight approximation of the rate coverage as depicted in Fig. \ref{fig:rc_1}. We will now study the impact of selection bias and node densities on the rate coverage of the typical receiver.

\begin{figure}
	\begin{minipage}{.45\textwidth}
		\centering
		\includegraphics[width=\textwidth]{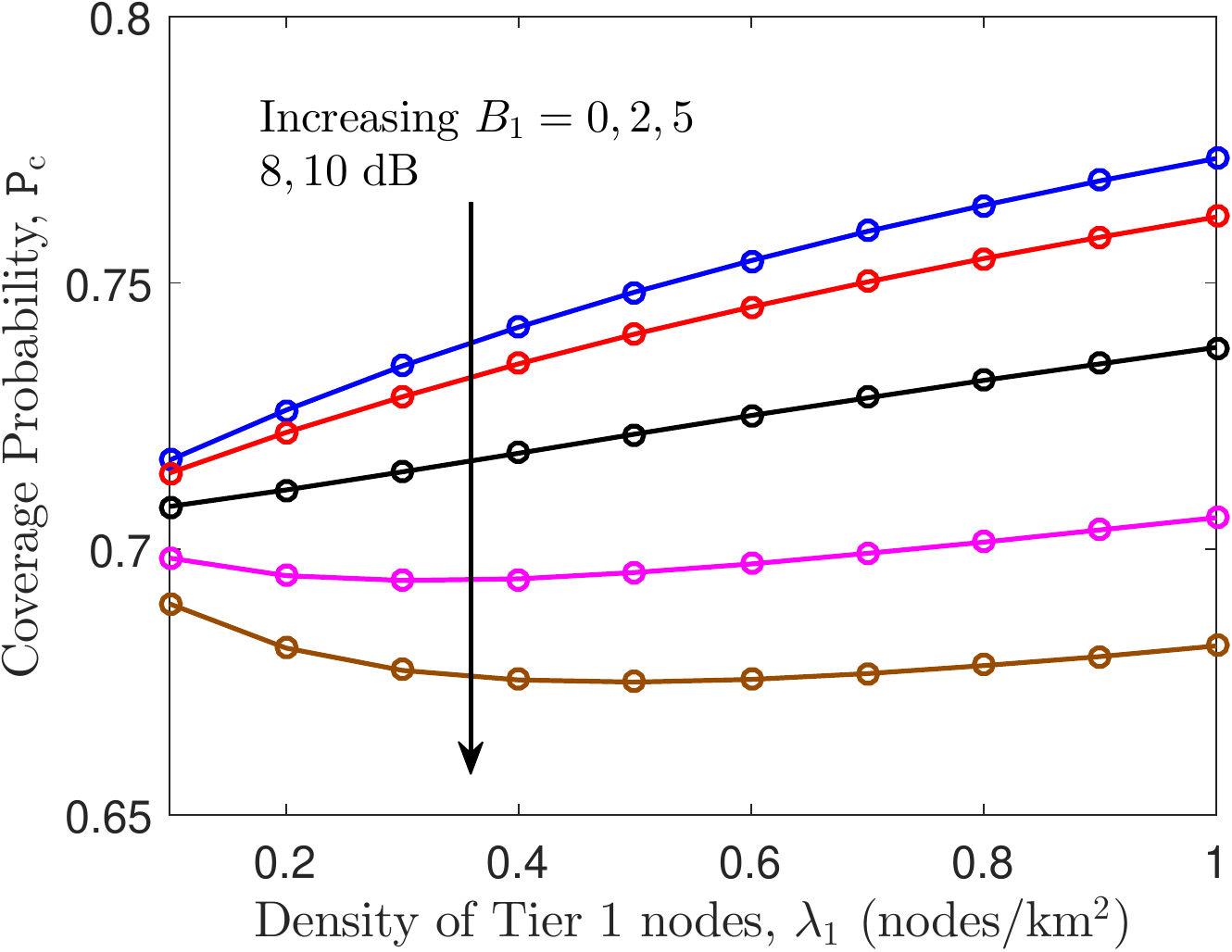}
		\caption{Coverage probability of the typical receiver as a function of the density of tier 1 nodes ($\mu_l$ = 5 km$^{-1}$, $\lambda_2 = 5$ nodes/km, $B_2 = 0 $ dB, $P_1 = 43$ dBm, $P_2 = 23$ dBm, $[\sigma_1\ \sigma_{20}\ \sigma_{21} ] = [ 4\ 2\ 4]$ dB, and $\beta = 0$ dB).}
		\label{fig:pc_3}
	\end{minipage}
\hfill
	\begin{minipage}{.45\textwidth}
		\centering
		\includegraphics[width=\textwidth]{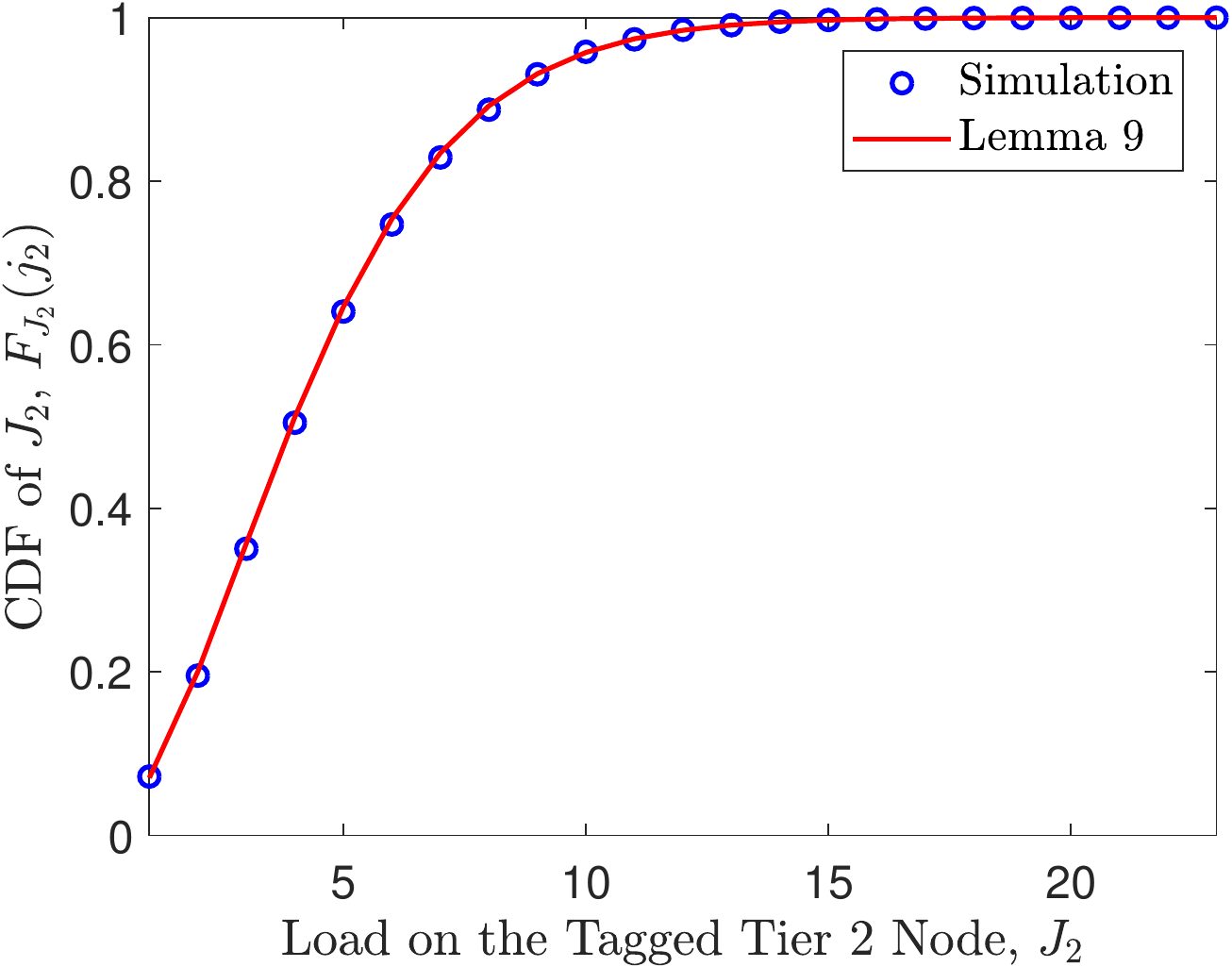}
		\caption{CDF of the load on the tagged tier 2 node ($\mu_l$ = 10 km$^{-1}$, $\lambda_1 = 0.5$ nodes/km$^2$, $\lambda_2 = 4$ nodes/km, $B_1 = B_2 =0 $ dB, $P_1 = 40$ dBm, $P_2 = 23$ dBm, and $[\sigma_1\ \sigma_{20}\ \sigma_{21} ] = [ 4\ 2\ 4]$ dB).}
		\label{fig:loadcomp}
	\end{minipage}

\end{figure}

 {\em Impact of selection bias.} We plot the rate coverage as a function of selection bias factor of tier 1 nodes $B_1$ for different values of node densities, as shown in Fig. \ref{fig:rc_2}. We observe that the rate coverage initially increases with $B_1$ and then degrades beyond a certain value, thereby yielding an optimal bias factor that maximizes the rate coverage of the typical receiver for a given set of node densities. Also, the optimal bias factor decreases as the density of tier 1 nodes increases. 
 
{\em Impact of node densities.} From Figs. \ref{fig:rc_2} and \ref{fig:rc_3}, we observe that the rate coverage improves as the density of tier 1 and tier 2 nodes increases when there is no selection bias. This is because of the reduced load on the tagged nodes as the density of nodes increases while the user density remains unchanged. However, we notice that this trend does not hold for higher values of selection bias $B_1$, as shown in Fig. \ref{fig:rc_2}. This is because of the downward trend in coverage probability for higher values of $B_1$, as shown in Figs. \ref{fig:pc_2} and \ref{fig:pc_3}. 

\begin{figure}
		\begin{minipage}{.45\textwidth}
		\centering
		\includegraphics[width=\textwidth]{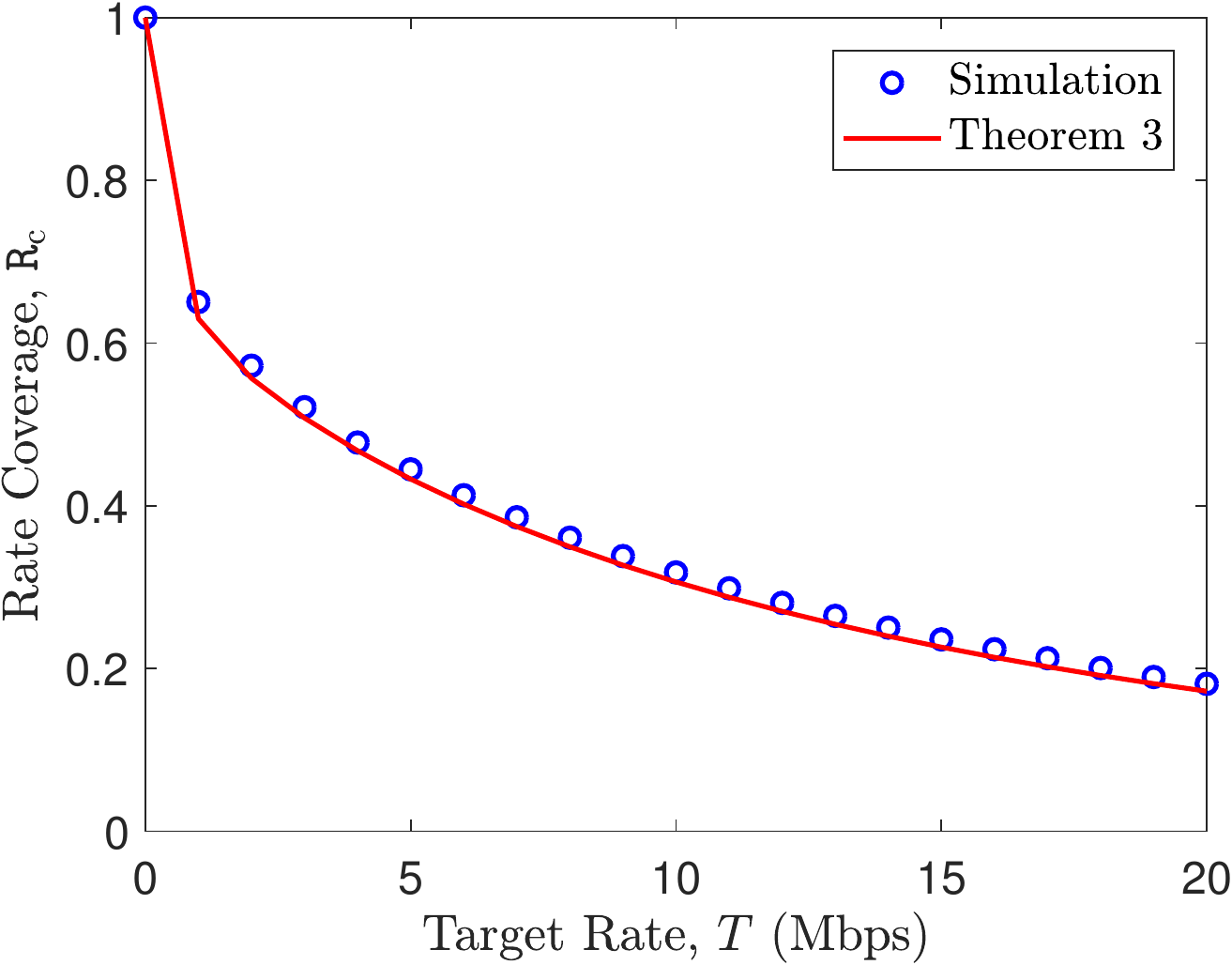}
		\caption{Rate coverage of the typical receiver as a function of target rate ($\mu_l$ = 10 km$^{-1}$, $\lambda_1 = 0.5$ nodes/km$^2$, $\lambda_2 = 4$ nodes/km, $B_1 = B_2 = 0 $ dB, $P_1 = 40$ dBm, $P_2 = 23$ dBm and $[\sigma_1\ \sigma_{20}\ \sigma_{21} ] = [ 4\ 2\ 4]$ dB).}
		\label{fig:rc_1}
	\end{minipage}
\hfill
	\begin{minipage}{.45\textwidth}
		\centering
		\includegraphics[width=\textwidth]{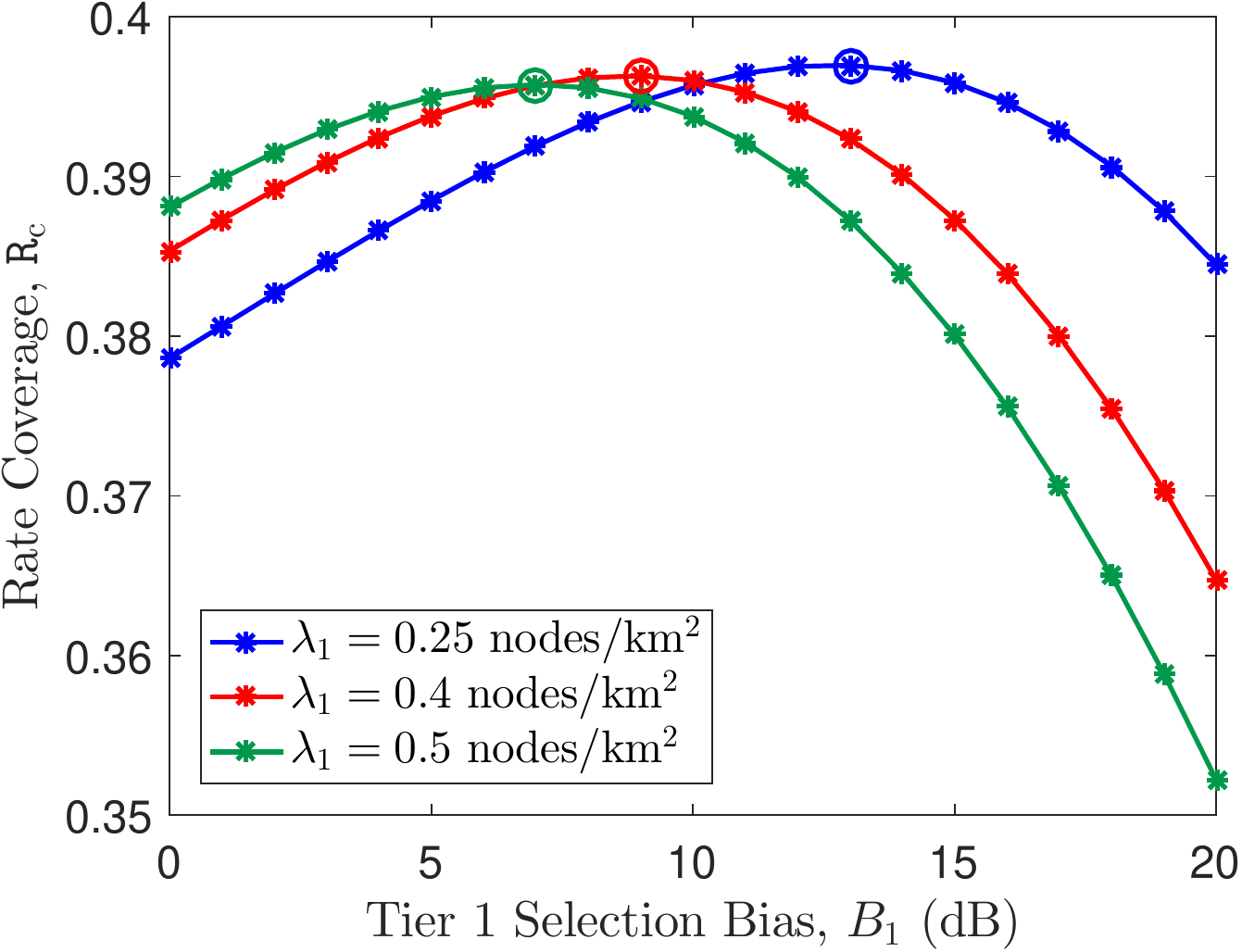}
		\caption{Rate coverage of the typical receiver as a function of selection bias $B_1$ ($\mu_l$ = 5 km$^{-1}$, $\lambda_2 = 5$ nodes/km, $B_2 = 0 $ dB, $P_1 = 43$ dBm, $P_2 = 23$ dBm, $[\sigma_1\ \sigma_{20}\ \sigma_{21} ] = [ 4\ 2\ 4]$ dB, and $T = 10$ Mbps).}
		\label{fig:rc_2}
	\end{minipage}

\end{figure}
\begin{figure}
	\centering 
\begin{minipage}{.45\textwidth}
	\centering
	\includegraphics[width=\textwidth]{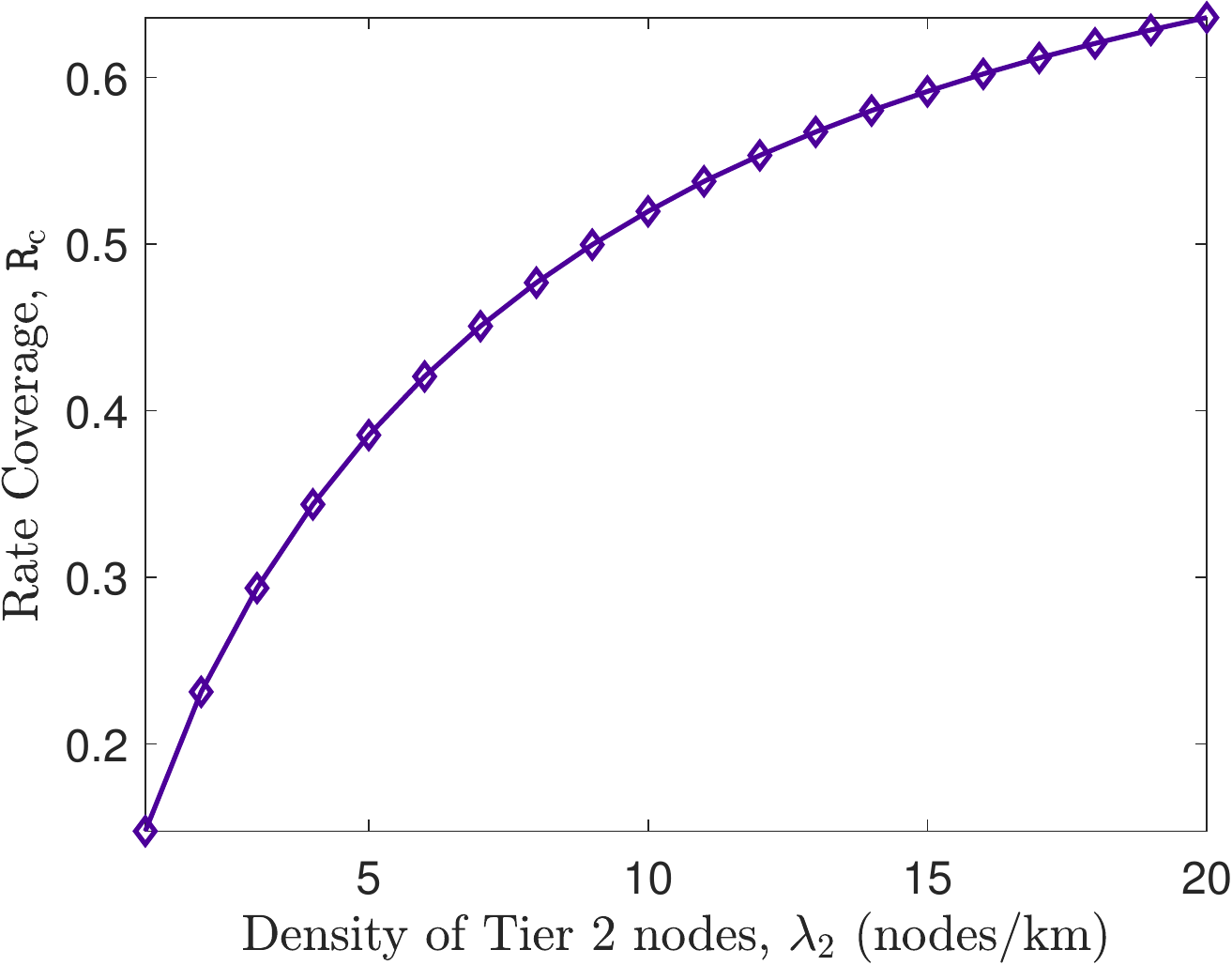}
	\caption{Rate coverage of the typical receiver as a function of the density of tier 2 nodes ($\mu_l$ = 5 km$^{-1}$, $\lambda_1 = 0.5$ nodes/km$^2$, $B_1 = B_2 = 0 $ dB, $P_1 = 43$ dBm, $P_2 = 23$ dBm, $[\sigma_1\ \sigma_{20}\ \sigma_{21} ] = [ 4\ 2\ 4]$ dB, and $T = 10$ Mbps).}
	\label{fig:rc_3}
\end{minipage}
\end{figure}
\subsection{Design Insights}
We will now provide some key system-level insights based on the trends observed in coverage probability and rate coverage. From Fig. \ref{fig:rc_2}, we observe that the rate coverage can be improved by increasing the density of tier 1 nodes. However, deploying more tier 1 nodes may not always be an option because of cost and site acquisition constraints. So, an alternate solution to improve the performance without additional deployment of nodes is to adjust the selection bias. For instance, we can observe that the rate coverage obtained by increasing the node density from $\lambda_1 = 0.25$ nodes/km$^2$ to $\lambda_1 = 0.5$ nodes/km$^2$ without any bias can also be achieved by increasing the selection bias to $B_1 = 5$ dB while keeping the node density unchanged. Unlike tier 1 nodes, we observe that an increase in the density of tier 2 nodes has a contrasting effect on coverage probability and rate coverage when there is no bias, as depicted in Figs. \ref{fig:pc_2} and \ref{fig:rc_3}. Hence, it is important to consider the trade-off between these two metrics in the deployment of RSUs.

\section{Conclusion}
In this paper, we have presented the coverage and rate analysis of a C-V2X network in the presence of shadowing. We have modeled the locations of vehicular nodes and RSUs as Cox processes driven by PLP and the locations of cellular MBSs as a 2D PPP. Assuming a fixed selection bias and maximum average received power based association, we computed the probability with which a typical receiver connects to another vehicular node or an RSU and a cellular MBS. Inspired by the asymptotic behavior of the Cox process, we approximated the Cox process of vehicular nodes and RSUs by a 2D PPP to characterize the interference from those nodes. We then derived the expression for SIR-based coverage probability in terms of the Laplace transform of interference power distribution. Further, we computed the rate coverage of the typical receiver by characterizing the load on the serving nodes. 
We have also provided several design insights based on the trends observed in the coverage probability and rate coverage as a function of network parameters. We have observed that the performance gain obtained by the densification of MBSs can be equivalently achieved by choosing an appropriate selection bias without additional deployment of infrastructure.

This work has numerous extensions. First of all, the proposed approximation of the spatial model based on the asymptotic behavior of the Cox process could be applied to lend tractability to similar analyses which may not otherwise be possible. For instance, the approximate spatial model and some of the intermediate results presented in this paper could be useful in {studying other metrics, such as latency and packet reception rate.} Another meaningful extension of this work could be to study the handover rate of moving vehicular users from not just one MBS to another but also from an RSU or a vehicular node to a MBS and vice versa. Further, there is scope for improving the system model by considering correlated shadowing where the nodes in close proximity experience similar loss in signal power due to blockages from the same buildings. Hence, it is worthwhile to develop generative blockage models that imitate the propagation of signals in a physical environment \cite{bacczhang, vishnuJ3}.

\appendix
\subsection{Proof of Lemma \ref{lem:void}} \label{app:void}
The void probability can be computed as
\begin{align*}
\P&\big(N_p (A) = 0\big) = \nbbE_{\Psi_\ell} \left[  \P\left( N_p \left(\cup_{L_i \in \Psi_{\ell}} \{ L_i \cap A\} \right) = 0 | \Psi_\ell\right)  \right] 
\stackrel{(a)}{=} \nbbE_{\Psi_\ell} \left[  \prod_{L_i \in \Psi_\ell}   \P \left( N_p ( L_i \cap A ) = 0 | \Psi_\ell \right)  \right]\\
&\stackrel{(b)}{=} \nbbE_{\Psi_\ell} \left[ \prod_{L_i \in \Psi_\ell} \exp \left( - \lambda_p \nu_1 \left(L_i \cap A\right)\right) \right] 
\stackrel{(c)}{=} \nbbE_{\Psi_\calC} \left[ \prod_{(\rho_i, \theta_i)\in \Psi_\calC} \exp \left( - \lambda_p \nu_1 \left(L_{(\rho_i, \theta_i)} \cap A\right)\right) \right]\\
&\stackrel{(d)}{=} \exp \left[- \lambda_\ell \int_0^{2\pi} \int_{\nbbR^+} \left[1 - \exp\left(  - \lambda_p \nu_1 \left(L_{(\rho, \theta)} \cap A\right) \right) \right] {\rm d} \rho  {\rm d} \theta \right] 
\end{align*}
where (a) follows from the independent distribution of points on the lines, (b) follows from the void probability of a 1D PPP, (c) follows from rewriting the expression in terms of the point process $\Psi_\calC$ in the representation space $\calC$, and (d) follows from the probability generating functional (PGFL) of the 2D PPP $\Psi_{\calC}$. 

\subsection{Proof of Theorem \ref{thm:convergencetoPPP}} \label{app:convergence}
In order to prove this result, by Choquet's theorem \cite{haenggi}, it is sufficient to show that the void probability of the Cox process converges to that of a 2D PPP. Therefore, we will now apply the limits $\lambda_\ell \to \infty$ and $\lambda_p \to 0$ on the expression for void probability derived in Lemma \ref{lem:void}. As the overall density of nodes in the network $\lambda_a = \pi \lambda_\ell \lambda_p$ remains constant, the application of the two limits $\lambda_\ell \to \infty$ and $\lambda_p \to 0$ can be simplified to a single limit by substituting $\lambda_p = \frac{\lambda_a}{\pi \lambda_\ell}$ in the expression in \eqref{eq:p0}. Thus, the asymptotic void probability can be evaluated as
\begin{align*}
\lim_{\lambda_\ell \to \infty} 	\P\big(N_p (A) = 0\big)	&= \lim_{\lambda_\ell \to \infty}  \exp \left[- \lambda_\ell \int_{0}^{2\pi} \int_{\nbbR^+} \left[1 - \exp\left(  - \lambda_p \nu_1 \left(L_{(\rho, \theta)} \cap A\right) \right) \right] {\rm d} \rho  {\rm d} \theta \right]\\
&= \exp \left[\lim_{\lambda_\ell \to \infty} - \lambda_\ell \int_{0}^{2\pi} \int_{\nbbR^+} \left[1 - \exp\left(  - \frac{\lambda_a}{\pi \lambda_\ell} \nu_1 \left(L_{(\rho, \theta)} \cap A\right) \right) \right] {\rm d} \rho  {\rm d} \theta  \right] \\
&\stackrel{(a)}{=}  \exp \left[\lim_{\lambda_\ell \to \infty}  \lambda_\ell \int_{0}^{2\pi} \int_{\nbbR^+}  \sum_{k=1}^{\infty} \frac{\left(-\lambda_a  \nu_1 \left(L_{(\rho, \theta)} \cap A\right) \right)^k}{\left(\pi \lambda_\ell\right)^{k} k!} {\rm d} \rho  {\rm d} \theta \right]\\
&\stackrel{(b)}{=} \exp \Bigg[  \lim_{\lambda_\ell \to \infty} \lambda_\ell \int_{0}^{2\pi} \mspace{-12mu} \int_{\nbbR^+} \frac{\left(-\lambda_a  \nu_1 \left(L_{(\rho, \theta)} \cap A\right) \right)}{\pi \lambda_\ell} {\rm d} \rho  {\rm d} \theta \\ 
&\hspace{6em}+  \sum_{k=2}^{\infty} \int_{0}^{2\pi} \mspace{-12mu}  \int_{\nbbR^+}  \lim_{\lambda_\ell \to \infty}\frac{\left(-\lambda_a  \nu_1 \left(L_{(\rho, \theta)} \cap A\right) \right)^k}{\pi^k \lambda_\ell^{k-1} k!} {\rm d} \rho  {\rm d} \theta\Bigg]\\
&\stackrel{(c)}{=}    \exp \left[\lim_{\lambda_\ell \to \infty}  -  \frac{\lambda_a}{\pi \lambda_\ell}\int_{0}^{2\pi} \int_{\nbbR^+} \nu_1 \left(L_{(\rho, \theta)} \cap A\right)  \lambda_\ell {\rm d} \rho  {\rm d} \theta \right]\\
&\stackrel{(d)}{=} \exp \left[ \lim_{\lambda_\ell \to \infty} -  \frac{\lambda_a}{\pi \lambda_\ell} \nbbE \left[ \sum_{ (\rho,\theta) \in \Psi_\calC } \nu_1\left(L_{(\rho, \theta)} \cap A\right) \right] \right]\\
&\stackrel{(e)}{=} \exp \left[\lim_{\lambda_\ell \to \infty} -\frac{\lambda_a}{\mu_\ell} \mu_\ell \nu_2 (A) \right] = \exp\left(-\lambda_a \nu_2(A) \right),
\end{align*}
where (a) follows from the Taylor series expansion of exponential function, (b) follows from switching the order of integral and summation operations and applying Dominated Convergence Theorem (DCT) on the second term, (c) follows from the limit of the integrand in the second term which evaluates to $0$ for all $k\geq2$, (d) follows from the Campbell's theorem for sums over stationary point processes, and (e) follows from the definition of line density of line processes, where $\nu_2(A)$ is the two-dimensional Lebesgue measure (area) of the region $A$.

\subsection{Proof of Lemma \ref{lem:pe1}} \label{app:pe1}

The typical receiver connects to a tier 1 node if the average biased received power from the closest tier 1 node exceeds the average biased received power from the closest tier 2 node on the typical line. Thus, the probability of occurrence of the event $\calE_1$ can be computed as
\begin{align*}
\P (\calE_1) &=  \P \left( P_1 B_1 G_1 R_1^{-\alpha} > P_2 B_2 G_2  R_2^{-\alpha}\right) 
\stackrel{(a)}{=} \nbbE_{R_2} \left[ \P \left( R_1 < \zeta_{21}^{-\frac{1}{\alpha}} r_2 | R_2\right) \right] \\
&=\int_0^{\infty} F_{R_1}\left(\zeta_{21}^{-\frac{1}{\alpha}} r_2 \right)	f_{R_2} (r_2) {\rm d}r_2\stackrel{(b)}{=} 1 - \int_0^{\infty} \exp\left[ -\lambda_1^{(e)} \pi \zeta_{21}^{-\frac{2}{\alpha}} r_2^2  \right] 2 \lambda_2^{(e)} \exp \left(- 2 \lambda_2^{(e)} r_2\right){\rm d}r_2 \\
&= 1- \lambda_2^{(e)} \sqrt{\frac{1}{\lambda_1^{(e)}  \zeta_{21}^{-\frac{2}{\alpha}}}} 
\exp \left[ \frac{(\lambda_2^{(e)})^2}{\lambda_1^{(e)} \pi \zeta_{21}^{-\frac{2}{\alpha}}}\right] 
\erfc \left(\frac{\lambda_2^{(e)}}{\sqrt{\lambda_1^{(e)} \pi \zeta_{21}^{-\frac{2}{\alpha}}}}\right),
\end{align*}
where (a) follows from substituting $\zeta_{21} = \frac{P_2 B_2 G_2}{P_1 B_1 G_1}$ and conditioning on $R_2$ , and (b) follows from substituting the expressions for $F_{R_1}(\cdot)$ and $f_{R_2}(\cdot)$ from \eqref{eq:cdfr1} and \eqref{eq:pdfr2} in the previous step. Since the events $\calE_1$ and $\calE_2$ are complementary,  $\P (\calE_2)  = 1- \P(\calE_1) $.

\subsection{Proof of Lemma \ref{lem:lIe1}} \label{app:lIe1}
	As the aggregate interference at the typical receiver can be decomposed into three independent components $I_{1}$, $I_{20}$, and $I_{21}$, the conditional Laplace transform of aggregate interference can be computed as the product of the conditional Laplace transforms of interference of these individual components, i.e.,
\begin{align}\label{eq:lIprod}
\calL_I(s|R, \calE_1) = \calL_{I_{1}}(s|R, \calE_1) \calL_{I_{20}}(s|R, \calE_1) \calL_{I_{21}}(s|R, \calE_1).
\end{align}
First, we will consider the interference from the tier 1 nodes $I_1$. Recall that the beamforming gain of an interfering tier 1 node is $G_1$ with probability $q_c$ and $g_1$ with probability $(1-q_c)$. So, we can partition the tier 1 interfering nodes into two independent 2D PPPs $\Phi_{11}^{(e)}$ and $\Phi_{12}^{(e)}$ with densities $q_c \lambda_1^{(e)}$ and $(1- q_c)\lambda_1^{(e)}$ and with beamforming gains $G_1$ and $g_1$, respectively. Thus, the Laplace transform of interference from the tier 1 nodes can be computed as
\begin{align}\notag
\calL_{I_{1}}&(s| R, \calE_1) = \nbbE\left[\exp(-s I_1)\right] = \nbbE\Bigg[ \exp\Bigg(-s \sum\limits_{\scalebox{1}{${\mathclap{\substack{ \nrmy \in \Phi_{11}^{(e)} \setminus b(o,R) } } }$}}  P_1 G_1 H_{1}  \|\nrmy\|^{-\alpha}  -s \sum\limits_{\scalebox{1}{${\mathclap{\substack{ \nrmy \in \Phi_{12}^{(e)} \setminus b(o,R) } } }$}}  P_1 g_1 H_{1}  \|\nrmy\|^{-\alpha}\Bigg)\Bigg]\\ \notag 
&\stackrel{(a)}{=} \nbbE_{\Phi_{11}^{(e)} }\nbbE_{H_1} \Bigg[ \prod_{ \nrmy \in \Phi_{11}^{(e)} \setminus b(o,R) }  e^{-s   P_1 G_1 H_{1}  \|\nrmy\|^{-\alpha} } \Bigg] 
\nbbE_{\Phi_{12}^{(e)} }\nbbE_{H_1} \Bigg[ \prod_{ \nrmy \in \Phi_{12}^{(e)} \setminus b(o,R) } e^{-s   P_1 g_1 H_{1}  \|\nrmy\|^{-\alpha} } \Bigg]\\ \notag 
&\stackrel{(b)}{=} \nbbE_{\Phi_{11}^{(e)} } \Bigg[ \prod_{\nrmy \in \Phi_{11}^{(e)} \setminus b(o,R) }  \left(1 + \frac{s P_1 G_1 \|\nrmy\|^{-\alpha}}{m_1}\right)^{-m_1}\Bigg] \nbbE_{\Phi_{11}^{(e)} } \Bigg[ \prod_{ \nrmy \in \Phi_{12}^{(e)} \setminus b(o,R) }  \left(1 + \frac{s P_1 g_1 \|\nrmy\|^{-\alpha}}{m_1}\right)^{-m_1}\Bigg] \\ \notag 
&\stackrel{(c)}{=} \exp \Bigg[ -2 \pi q_c \lambda_1^{(e)} \int_r^{\infty} 1 - \left(1 + \frac{s P_1 G_1 y^{-\alpha}}{m_1}\right)^{-m_1} y {\rm d} y \\ \label{eq:lI1e1}
& \hspace{11em} - 2 \pi (1- q_c) \lambda_1^{(e)} \int_r^{\infty} 1 -\left(1 + \frac{s P_1 g_1 y^{-\alpha}}{m_1}\right)^{-m_1} y {\rm d}y \Bigg],
\end{align}
where (a) follows from the independence of $\Phi_{11}^{(e)}$ and $\Phi_{12}^{(e)}$, (b) follows from the Nakagami-$m$ fading assumption, and (c) follows from substituting $y = \| \nrmy \|$ and the PGFL of a 2D PPP. 
The Laplace transform of interference from tier 2 nodes located on the typical line conditioned on $R$ and $\calE_1$ can be computed as $\calL_{I_{20}}(s|R, \calE_1)=$
\begin{align}\notag 
&\nbbE\Bigg[ \exp \Bigg( -s \mspace{-25mu} \sum_{ \nrmy \in \Xi_{L_0}^{(e)} \setminus [-\zeta_{21}^{\frac{1}{\alpha}}r, \zeta_{21}^{\frac{1}{\alpha}}r  ] } \mspace{-30mu} P_2 G_2 H_{20} \| \nrmy \|^{-\alpha} \Bigg)\Bigg] = \nbbE_{\Xi_{L_0}^{(e)}} \nbbE_{H_{20}} \Bigg[ \mspace{-5mu} \prod_{ \nrmy \in \Xi_{L_0}^{(e)} \setminus [-\zeta_{21}^{\frac{1}{\alpha}}r, \zeta_{21}^{\frac{1}{\alpha}}r  ] } \mspace{-30mu} e^{ -s P_2 G_2 H_{20} \|\nrmy\|^{-\alpha} } \Bigg] \\ \label{eq:lI20e1}
&\stackrel{(a)}{=} \exp \Bigg[- 2  \lambda_2^{(e)} \int_{\zeta_{21}^{\frac{1}{\alpha}} r}^{\infty} 1 - \bigg(1 + \frac{s P_2 G_2 y^{-\alpha }}{m_{20}}\bigg)^{-m_{20}} {\rm d} y \Bigg],
\end{align}
where (a) simply follows from the Nakagami-$m$ fading assumption and the PGFL of 1D PPP.
Under Assumption \ref{assum:ppp}, the Laplace transform of $I_{21}$ conditioned on $R$ and $\calE_1$ can be computed as
\begin{align}\notag
\calL_{I_{21}}(s|R, \calE_1) &= \nbbE \Bigg[\exp \Bigg( -s \sum_{\nrmy \in \Phi_2^{(a)}} P_2 g_2 H_{21} \|\nrmy\|^{-\alpha} \Bigg)\Bigg] = \nbbE_{\Phi_{2}^{(a)} } \nbbE_{H_{21}} \Bigg[ \prod_{\nrmy \in \Phi_2^{(a)} }   e^{-s P_2 g_2 H_{21} \| \nrmy \|^{-\alpha} }\Bigg] \\\label{eq:lI21e1}
&\stackrel{(a)}{=}\exp \Bigg[- 2 \pi \lambda_2^{(a)} \int_{0}^{\infty} 1 - \bigg(1 + \frac{s P_2 g_2 y^{-\alpha }}{m_{21}}\bigg)^{-m_{21}} y {\rm d} y \Bigg],
\end{align}
where (a) follows from the Nakagami-$m$ fading assumption and PGFL of a 2D PPP.

Substituting \eqref{eq:lI1e1}, \eqref{eq:lI20e1}, and \eqref{eq:lI21e1} in \eqref{eq:lIprod}, we obtain the final expression for the conditional Laplace transform of the aggregate interference power distribution.

\subsection{Proof of Lemma \ref{lem:pdfztyp}} \label{app:pdfztyp}
	The total length of the typical cell is the sum of the distances to the farthest points on the line on either side of the typical node such that any user closer than that point would connect to the typical node. We denote the farthest points of the typical cell by ${\rm z}_0$ and ${\rm z}_1$ and the distances to these points from the typical node by $Z_0$ and $Z_1$, respectively. Thus, the length of the typical cell is $Z_{\rm typ} = Z_0 + Z_1$. Without loss of generality, we assume that ${\rm z_0}$ is to the right of ${\nrmx_{\rm typ}}$ and we will first focus on the distance $Z_0$.

The CDF of $Z_0$ can be computed as 
\begin{align*}
F_{Z_0}(z_0) &= 1 - \P (Z_0 > z_0)= 1 - \P \big( N_2(L \cap b\left({\rm z_0}, z_0\right) )= 0 \big) \P \left( N_1 \left( b \left({\rm z_0},\zeta_{21}^{-\frac{1}{\alpha}} z_0 \right)  \right) = 0\right)\\
& = 1 - \exp\left[- 2 \lambda_2^{(e)} z_0 - \lambda_1^{(e)} \pi\zeta_{21}^{-\frac{2}{\alpha}} z_0^2\right],
\end{align*} 
where $N_1(\cdot)$ and $N_2(\cdot)$ denote number of tier 1 and tier 2 nodes, respectively, and $b(c, d)$ denotes a ball of radius $d$ centered at $c$. As the random variables $Z_0$ and $Z_1$ are not independent, we will now compute the CDF of $Z_1$ conditioned on $Z_0$. Before we proceed with that derivation, it is important to note that the boundary of the typical cell ${\rm z_0}$ could have been determined by an adjacent tier 2 node on the same line or a tier 1 node. In the former case, conditioning on $Z_0$ implies that there is a tier 2 node on the same line at a distance $2 z_0$ from ${\rm x_{typ}}$. On the other hand, if the boundary was determined by a tier 1 node, conditioning on $Z_0$ also means that there exists a tier 1 node on the edge of the disc $b({\rm z_0}, k z_0)$, where $k=\zeta_{21}^{-\frac{1}{\alpha}}$. Therefore, in order to exactly characterize the length of the typical cell, we need to distinguish these two cases in the derivation of the conditional distribution of $Z_1$. However, this would result in multiple sub-cases and hence complicate the analysis. Therefore, in the interest of tractability, we do not distinguish these two cases in our analysis. As will be shown in the next section, this does not affect the accuracy of our final results. Thus, the CDF of $Z_1$ conditioned on $Z_0$ can be computed as 	
\begin{figure}
	\begin{minipage}{.28\textwidth}
		\centering
		\includegraphics[width=\textwidth]{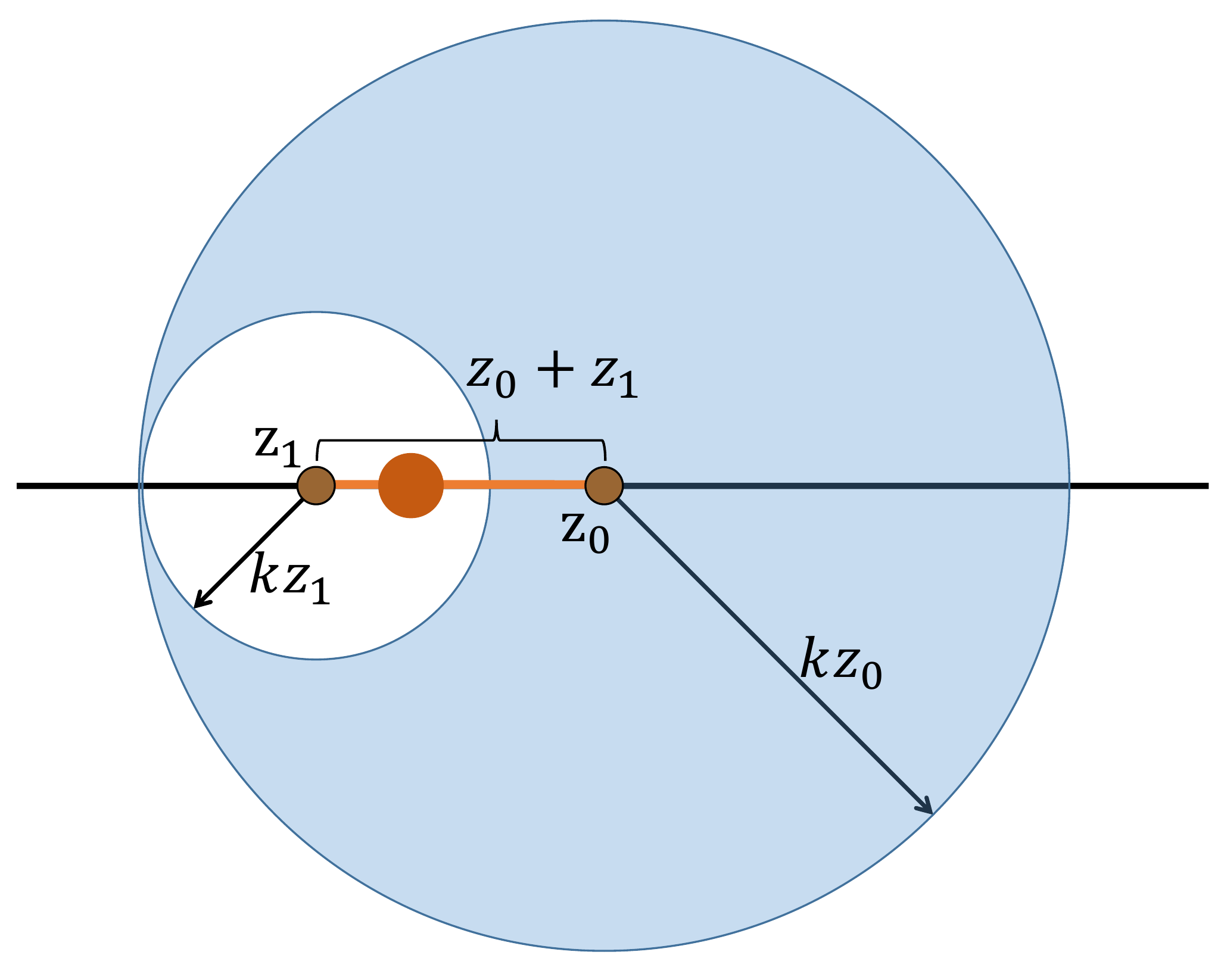}
		{(i)}
		\label{fig:casea}
	\end{minipage}
	\hfill
	\begin{minipage}{.41\textwidth}
		\centering
		\includegraphics[width=\textwidth]{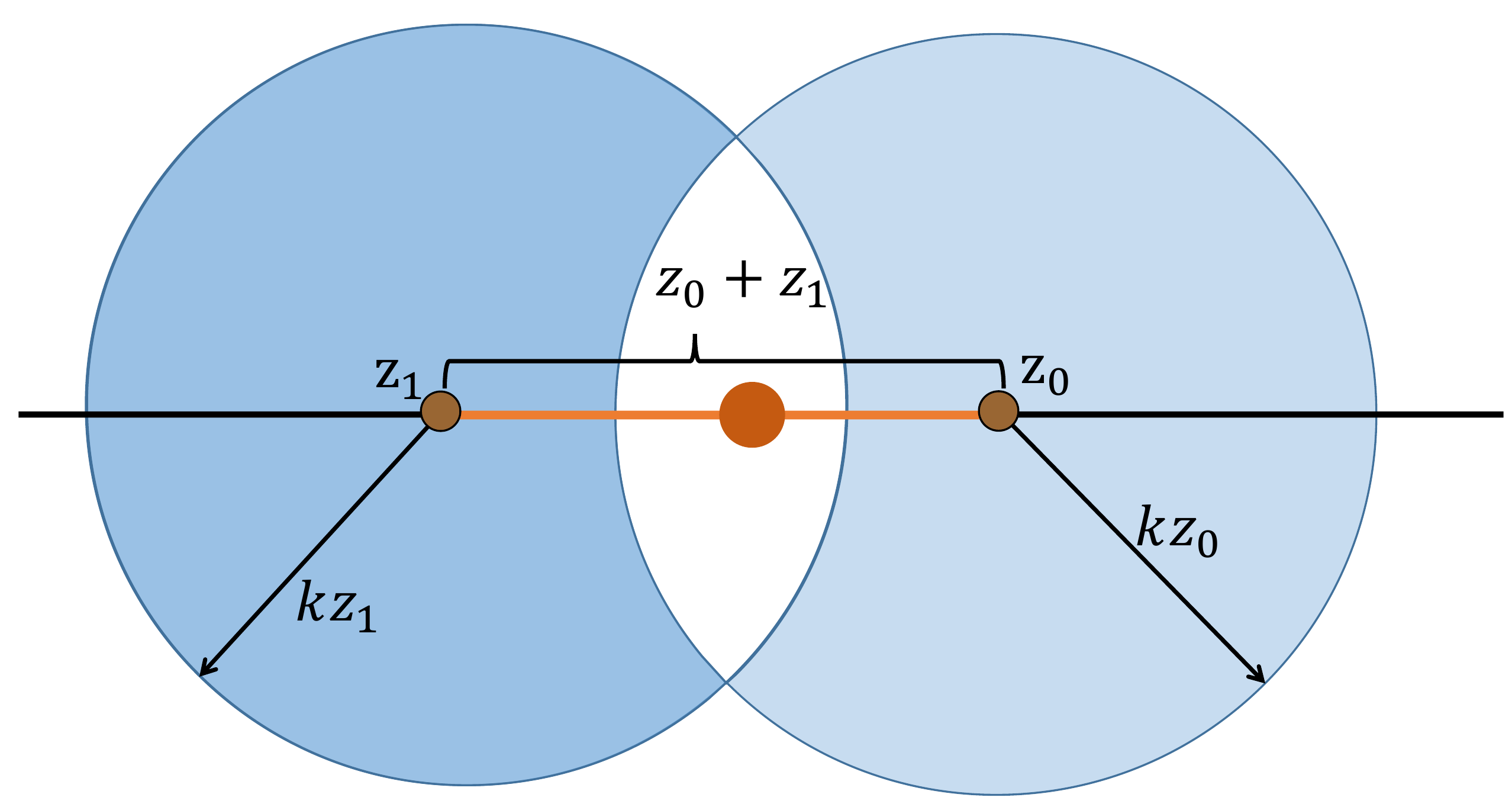}
		{\text{\newline}  (ii)}
	\end{minipage}
	\hfill
	\begin{minipage}{.28\textwidth}
		\centering
		\includegraphics[width=\textwidth]{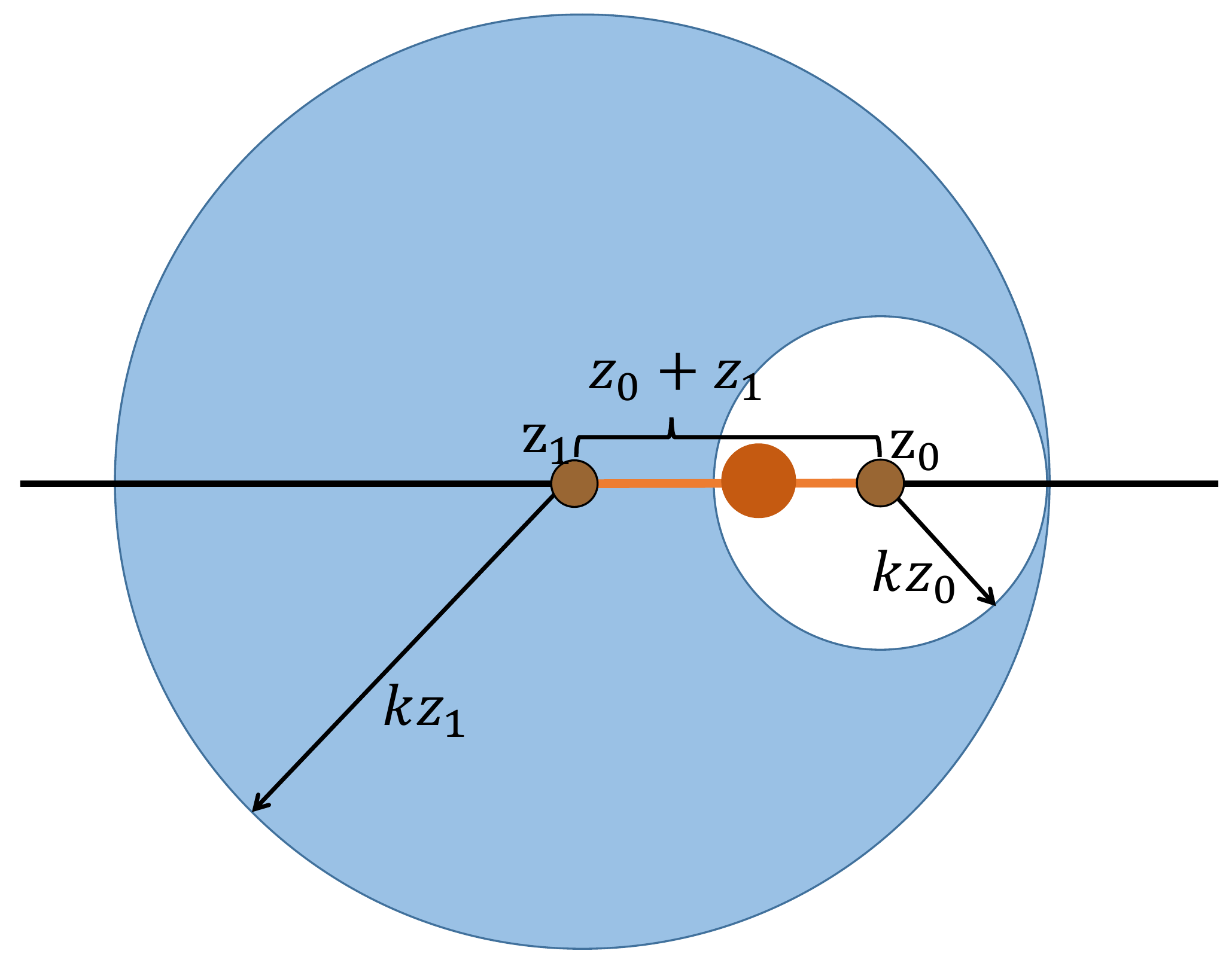}
		{(iii)}
	\end{minipage} 	
	\caption{An illustration of the three cases: (i) $0<z_1 < \frac{k-1}{k+1}z_0$, (ii) $\frac{k-1}{k+1}z_0 < z_1 < \frac{k+1}{k-1}z_0$, and (iii) $\frac{k+1}{k-1}z_0 < z_1 < \infty$.}
	\label{fig:cases}
\end{figure}
\begin{align}\notag
F_{Z_1}(z_1| z_0) &= 1 - \P (Z_1 > z_1 | Z_0) \\\notag
&\approx1 - \P \left(  \{ N_2(L \cap b\left({\rm z_1}, z_1\right) )= 0  \} \cap \{ N_1 \left( b( {\rm z_1}, k z_1 ) \right) = 0 \} \big| N_1 \left(b({\rm z_0}, k z_0) \right) = 0  \right) \\ \notag 
&= 1 - \P \left(  N_2((L \cap b\left({\rm z_1}, z_1\right) )= 0  \right) \P \left( N_1 \left( b( {\rm z_1}, k z_1 ) \setminus  b({\rm z_0}, k z_0)\right) = 0 \right)\\ \label{eq:cdfz1_step1}
&= 1- \exp(-2 \lambda_{2}^{(e)} z_1) \exp \left(- \lambda_1^{(e)} \nu_2( b( {\rm z_1}, k z_1 ) \setminus  b({\rm z_0}, k z_0) ) \right).
\end{align}
The area of the region $ b( {\rm z_1}, k z_1 ) \setminus  b({\rm z_0}, k z_0)$ is determined by the relative ranges of $z_0$ and $z_1$, as shown in Fig. \ref{fig:cases}. Hence, we obtain a piecewise function for this area which is given by
\begin{align}\notag
&\nu_2 \Big(  b( {\rm z_1}, k z_1 ) \setminus  b({\rm z_0}, k z_0) \Big) = \begin{dcases}
\gamma_{2,1} (z_0, z_1),& 0<z_1 < \frac{k-1}{k+1}z_0, \\
\gamma_{2,2} (z_0, z_1), &\frac{k-1}{k+1}z_0 < z_1 < \frac{k+1}{k-1}z_0\\
\gamma_{2,3} (z_0, z_1), & \frac{k+1}{k-1}z_0 < z_1 < \infty
\end{dcases} \\ \label{eq:area_pw}
&= \begin{dcases}
0 & 0<z_1 < \frac{k-1}{k+1}z_0,\\ 
\pi k^2 z_1^2 - (k z_0)^2 (\theta -\frac{1}{2} \sin(2\theta) ) - (k z_1)^2 (\phi -\frac{1}{2} \sin(2\phi)), \quad &\frac{k-1}{k+1}z_0 < z_1 < \frac{k+1}{k-1}z_0 \\
\pi (kz_1)^2 - \pi (kz_0)^2 , & \frac{k+1}{k-1}z_0 < z_1 < \infty
\end{dcases},
\intertext{where }
&\notag \theta = \arccos \Bigg(\frac{(z_0 + z_1)^2 + (k z_0)^2 - (k z_1)^2}{2 k z_0 ( z_0 + z_1)}\Bigg), \textrm{and } \phi = \arccos \Bigg( \frac{(z_0 + z_1)^2 - (k z_0)^2 + (k z_1)^2}{2 k z_1 ( z_0 + z_1)}\Bigg).
\end{align}
Substituting \eqref{eq:area_pw} in \eqref{eq:cdfz1_step1}, we obtain the conditional CDF of $Z_1$ as follows
\begin{align}
F_{Z_1} = \begin{dcases}
1 - e^{-2 \lambda_2^{(e)} z_1}, & 0<z_1 < \frac{k-1}{k+1}z_0, \\
1 - e^{-2 \lambda_2^{(e)} z_1 - \lambda_1^{(e)} \gamma_{2,2}(z_0, z_1) },  &\frac{k-1}{k+1}z_0 < z_1 < \frac{k+1}{k-1}z_0,\\
1 - e^{- 2  \lambda_2^{(e)} z_1 - \lambda_1^{(e)} \pi k^2 z_1^2 + \lambda_1^{(e)} \pi k^2 z_0^2 }, & \frac{k+1}{k-1}z_0 < z_1 < \infty.
\end{dcases}
\end{align}

Having determined the CDF of $Z_0$ and conditional CDF of $Z_1$, the CDF of the length of the typical cell can be computed as 
\begin{flalign}\notag 
F_{Z_{\rm typ}}(z) &= \P ( Z_0  + Z_1 < z) = \int_0^z \P (Z_1 < z- z_0 | Z_0) f_{Z_0}(z_0) {\rm d} z_0 = \int_0^z F_{Z_1}(z - z_0| z_0) f_{Z_0}(z_0) {\rm d}z_0 \\\notag 
&= \int_{\frac{k+1}{2k}z}^{z} F_{Z_1,1}(z - z_0) f_{Z_0}(z_0) {\rm d}z_0  + \int_{\frac{k-1}{2k}z}^{\frac{k+1}{2k}z} F_{Z_1,2}(z - z_0) f_{Z_0}(z_0) {\rm d}z_0  \\
& \hspace{18em} + \int_0^{\frac{k-1}{2k}z} F_{Z_1,3}(z - z_0) f_{Z_0}(z_0) {\rm d}z_0
\end{flalign}
The PDF of $Z_{\rm typ}$ can be obtained by taking the derivative of $F_{Z_{\rm typ}}(z)$ w.r.t. $z$.

\subsection{Proof of Lemma \ref{lem:barJ1}} \label{app:meanload}
The mean load on tagged tier 1 node can be obtained by determining the difference between the mean number of vehicular users inside $\calV_{\rm tag}$ and the mean number of vehicular users served by tier 2 nodes inside $\calV_{\rm tag}$. Thus, the mean load can be computed as
\begin{align}\label{eq:meanload_step1}
\nbbE[ J_1 ] = 1 + \nbbE\left[ N_r(\calV_{\rm tag})  \right] - \nbbE \left[ N_2(\calV_{\rm tag})  \right] \nbbE\left[J_{2}' \right],
\end{align}
where $N_r(\cdot)$ represents the number of receiving vehicular user nodes and $J_{2}'$ denotes the load on a typical tier 2 node.
We will now determine each term in the RHS of the above equation. The first term, which represents the average number of vehicles in $\calV_{\rm tag}$, is given by
\begin{align}\label{eq:meanload_term1}
\nbbE\left[ N_r(\calV_{\rm tag}) \right]  = \lambda_r \nbbE \left[ \sum_{L_i \in \Phi_{l_0}} \nu_1( L_i \cap \calV_{\rm tag} ) \right] = \left( \frac{1.28 \pi \lambda_l  }{\lambda_1^{(e)}}+ \frac{3.216}{\pi \sqrt{\lambda_1^{(e)}}}\right) \lambda_r.
\end{align}
The proof of this result is given in \cite[Theorem 2]{bartek} and is hence skipped.
Similarly, the second term in \eqref{eq:meanload_step1} which represents the average number of tier 2 nodes in $\calV_{\rm tag}$ is given by 
\begin{align}\label{eq:meanload_term2}
\nbbE \left[ N_2(\calV_{\rm tag})  \right] =\left( \frac{1.28 \pi \lambda_l  }{\lambda_1^{(e)}}+ \frac{3.216}{\pi \sqrt{\lambda_1^{(e)}}}\right) \lambda_2^{(e)}. 
\end{align}

The third term in \eqref{eq:meanload_step1}, which corresponds to the mean load on a typical tier 2 node, can be easily computed using distribution of the length of the typical cell as 
\begin{align}\label{eq:meanload_term3}
\nbbE \left[ J_2' \right] = \lambda_r \nbbE[ Z_{\rm typ}] = \lambda_r \int_0^{\infty} z f_{Z_{\rm typ}} (z) {\rm d}z,
\end{align}
where $f_{Z_{\rm typ}} (z)$ is given in Lemma \ref{lem:pdfztyp}. Substituting \eqref{eq:meanload_term1}, \eqref{eq:meanload_term2}, and \eqref{eq:meanload_term3} in \eqref{eq:meanload_step1}, we obtain the final expression for the mean load on the tagged tier 1 node.

{ \setstretch{1.2}
\bibliographystyle{IEEEtran}
\bibliography{j4_0.7.bbl}

\begin{thebibliography}{10}
\providecommand{\url}[1]{#1}
\csname url@samestyle\endcsname
\providecommand{\newblock}{\relax}
\providecommand{\bibinfo}[2]{#2}
\providecommand{\BIBentrySTDinterwordspacing}{\spaceskip=0pt\relax}
\providecommand{\BIBentryALTinterwordstretchfactor}{4}
\providecommand{\BIBentryALTinterwordspacing}{\spaceskip=\fontdimen2\font plus
\BIBentryALTinterwordstretchfactor\fontdimen3\font minus
  \fontdimen4\font\relax}
\providecommand{\BIBforeignlanguage}[2]{{%
\expandafter\ifx\csname l@#1\endcsname\relax
\typeout{** WARNING: IEEEtran.bst: No hyphenation pattern has been}%
\typeout{** loaded for the language `#1'. Using the pattern for}%
\typeout{** the default language instead.}%
\else
\language=\csname l@#1\endcsname
\fi
#2}}
\providecommand{\BIBdecl}{\relax}
\BIBdecl

\bibitem{survey}
H.~Hartenstein and L.~P. Laberteaux, ``A tutorial survey on vehicular ad hoc
  networks,'' \emph{IEEE Commun. Magazine}, vol.~46, no.~6, pp. 164--171, Jun.
  2008.

\bibitem{traffic_safety}
S.~Biswas, R.~Tatchikou, and F.~Dion, ``Vehicle-to-vehicle wireless
  communication protocols for enhancing highway traffic safety,'' \emph{IEEE
  Commun. Magazine}, vol.~44, no.~1, pp. 74--82, Jan. 2006.

\bibitem{vc_its}
P.~Papadimitratos, A.~D.~L. Fortelle, K.~Evenssen, R.~Brignolo, and S.~Cosenza,
  ``Vehicular communication systems: Enabling technologies, applications, and
  future outlook on intelligent transportation,'' \emph{IEEE Commun. Magazine},
  vol.~47, no.~11, pp. 84--95, Nov. 2009.

\bibitem{usecases}
M.~Boban, A.~Kousaridas, K.~Manolakis, J.~Eichinger, and W.~Xu, ``Connected
  roads of the future: Use cases, requirements, and design considerations for
  vehicle-to-everything communications,'' \emph{IEEE Veh. Technology Magazine},
  vol.~13, no.~3, pp. 110--123, Sep. 2018.

\bibitem{3gpp}
{3GPP TR 36.885}, ``{Study on LTE-based V2X services},'' Jul. 2016.

\bibitem{vishnuJ2}
V.~V. Chetlur and H.~S. Dhillon, ``Coverage analysis of a vehicular network
  modeled as {Cox} process driven by {Poisson} line process,'' \emph{IEEE
  Trans. on Wireless Commun.}, vol.~17, no.~7, pp. 4401--4416, Jul. 2018.

\bibitem{bacc_plp}
F.~Baccelli, M.~Klein, M.~Lebourges, and S.~Zuyev, ``Stochastic geometry and
  architecture of communication networks,'' \emph{Telecommunication Systems},
  vol.~7, no.~1, pp. 209--227, Jun. 1997.

\bibitem{mit}
B.~Blaszczyszyn, P.~Muhlethaler, and Y.~Toor, ``Maximizing throughput of linear
  vehicular {Ad-hoc NETworks} ({VANETs}) -- a stochastic approach,'' in
  \emph{European Wireless Conf.}, May 2009, pp. 32--36.

\bibitem{busanelli}
S.~Busanelli, G.~Ferrari, and R.~Gruppini, ``Performance analysis of broadcast
  protocols in {VANETs} with {Poisson} vehicle distribution,'' in \emph{Intl.
  Conf. on ITS Telecommunications}, Aug 2011, pp. 133--138.

\bibitem{mabiala}
M.~Mabiala, A.~Busson, and V.~Veque, ``Inside {VANET}: Hybrid network
  dimensioning and routing protocol comparison,'' in \emph{Proc., IEEE Veh.
  Technology Conf.}, Apr. 2007, pp. 227--232.

\bibitem{Ni}
M.~Ni, M.~Hu, Z.~Wang, and Z.~Zhong, ``Packet reception probability of {VANETs}
  in urban intersecton scenario,'' in \emph{Intl. Conf. on Connected Vehicles
  and Expo}, Oct. 2015, pp. 124--125.

\bibitem{Steinmetz}
E.~Steinmetz, M.~Wildemeersch, T.~Q.~S. Quek, and H.~Wymeersch, ``A stochastic
  geometry model for vehicular communication near intersections,'' in
  \emph{Proc., IEEE Globecom Workshops}, Dec. 2015, pp. 1--6.

\bibitem{vishnuL1}
V.~V. Chetlur and H.~S. Dhillon, ``Success probability and area spectral
  efficiency of a {VANET} modeled as a {Cox} process,'' \emph{IEEE Wireless
  Commun. Letters}, vol.~7, no.~5, pp. 856--859, Oct. 2018.

\bibitem{multihop}
B.~Blaszczyszyn and P.~Muhlethaler, ``Random linear multihop relaying in a
  general field of interferers using spatial {Aloha},'' \emph{IEEE Trans. on
  Wireless Commun.}, vol.~14, no.~7, pp. 3700--3714, Jul. 2015.

\bibitem{robert}
Y.~Wang, K.~Venugopal, A.~F. Molisch, and R.~W. Heath, ``{MmWave}
  vehicle-to-infrastructure communication: Analysis of urban microcellular
  networks,'' \emph{IEEE Trans. on Veh. Technology}, vol.~67, no.~8, pp.
  7086--7100, Aug. 2018.

\bibitem{morlot}
F.~Morlot, ``A population model based on a {Poisson} line tessellation,'' in
  \emph{Proc., Modeling and Optimization in Mobile, Ad Hoc and Wireless
  Networks}, May 2012, pp. 337--342.

\bibitem{bartek}
A.~Chattopadhyay, B.~Blaszczyszyn, and E.~Altman, ``Two-tier cellular networks
  for throughput maximization of static and mobile users,'' \emph{IEEE Trans.
  on Wireless Commun.}, 2018, to appear.

\bibitem{dim5g}
J.~Rachad, R.~Nasri, and L.~Decreusefond, ``How to dimension {5G} network when
  users are distributed on roads modeled by {Poisson} line process,''
  \emph{arXiv preprint}, 2018, available online: arxiv.org/abs/1805.06637.

\bibitem{bacc_letter}
C.~Choi and F.~Baccelli, ``Poisson {Cox} point processes for vehicular
  networks,'' \emph{IEEE Trans. on Veh. Technology}, vol.~67, no.~10, pp.
  10\,160--10\,165, Oct. 2018.

\bibitem{volker}
F.~Voss, C.~Gloaguen, F.~Fleischer, and V.~Schmidt, ``Distributional properties
  of {Euclidean} distances in wireless networks involving road systems,''
  \emph{IEEE Journal on Sel. Areas in Commun.}, vol.~27, no.~7, pp. 1047--1055,
  Sep. 2009.

\bibitem{vishnuICC}
V.~V. Chetlur, S.~Guha, and H.~S. Dhillon, ``Characterization of {V2V} coverage
  in a network of roads modeled as {Poisson} line process,'' in \emph{Proc.,
  IEEE Intl. Conf. on Commun. (ICC)}, 2018, pp. 1--6.

\bibitem{baccchoi}
C.~Choi and F.~Baccelli, ``An analytical framework for coverage in cellular
  networks leveraging vehicles,'' \emph{IEEE Trans. on Commun.}, vol.~66,
  no.~10, pp. 4950--4964, Oct. 2018.

\bibitem{sguha}
S.~Guha, ``Cellular-assisted vehicular communications: A stochastic geometric
  approach,'' Master's thesis, Virginia Tech, 2016.

\bibitem{Sial}
M.~N. Sial, Y.~Deng, J.~Ahmed, A.~Nallanathan, and M.~Dohler, ``Stochastic
  geometry modeling of cellular {V2X} communication on shared uplink
  channels,'' \emph{arXiv preprint}, 2018, available online:
  arxiv.org/abs/1804.08409.

\bibitem{haenggi}
M.~Haenggi, \emph{Stochastic Geometry for Wireless Networks}.\hskip 1em plus
  0.5em minus 0.4em\relax Cambridge University Press, 2013.

\bibitem{Dhi_letter}
H.~S. Dhillon and J.~G. Andrews, ``Downlink rate distribution in heterogeneous
  cellular networks under generalized cell selection,'' \emph{IEEE Wireless
  Commun. Letters}, vol.~3, no.~1, pp. 42--45, Feb. 2014.

\bibitem{stoyan}
S.~N. Chiu, D.~Stoyan, W.~S. Kendall, and J.~Mecke, \emph{Stochastic geometry
  and its applications}.\hskip 1em plus 0.5em minus 0.4em\relax John Wiley \&
  Sons, 2013.

\bibitem{antenna_meas}
V.~Shivaldova, A.~Paier, D.~Smely, and C.~F. Mecklenbrauker, ``On roadside unit
  antenna measurements for vehicle-to-infrastructure communications,'' in
  \emph{Proc., IEEE PIMRC}, Sep. 2012, pp. 1295--1299.

\bibitem{patil}
G.~P. Patil, ``Weighted distributions,'' \emph{Wiley StatsRef: Statistics
  Reference Online}, 2014.

\bibitem{offlsshd}
S.~Singh, H.~S. Dhillon, and J.~G. Andrews, ``Offloading in heterogeneous
  networks: Modeling, analysis, and design insights,'' \emph{IEEE Trans. on
  Wireless Commun.}, vol.~12, no.~5, pp. 2484--2497, May 2013.

\bibitem{bacczhang}
F.~Baccelli and X.~Zhang, ``A correlated shadowing model for urban wireless
  networks,'' in \emph{Proc., IEEE INFOCOM}.\hskip 1em plus 0.5em minus
  0.4em\relax IEEE, 2015, pp. 801--809.

\bibitem{vishnuJ3}
V.~V. Chetlur, H.~S. Dhillon, and C.~P. Dettmann, ``Characterizing shortest
  paths in road systems modeled as {Manhattan} {Poisson} line processes,''
  \emph{arXiv preprint}, 2018, available online: arxiv.org/abs/1811.11332.

\end{thebibliography}
}
\end{document}